\documentclass[aps,prd,superscriptaddress,nofootinbib,tighten,preprint]{revtex4}
\usepackage{float} 
\usepackage[utf8]{inputenc}
\usepackage[english]{babel}
\usepackage[utf8]{inputenc}
\usepackage[T1]{fontenc}
\usepackage{amsmath}
\usepackage{subfigure}
\usepackage{amsfonts}
\usepackage{amssymb}
\usepackage{graphicx,color,changes}
\newcommand{\be}{\begin{equation}}
\newcommand{\ee}{\end{equation}}
\newcommand{\bea}{\begin{eqnarray}}
\newcommand{\eea}{\end{eqnarray}}

\catcode`@=12

\begin{document}

\title{Decaying vector dark matter with low reheating temperature for KM3NeT signal and its impact on gravitational waves}

\author{Sarif Khan}
\email{sarifkhan@cau.ac.kr}
\affiliation{Department of Physics, Chung-Ang University, Seoul 06974, Korea.}
\author{Jongkuk Kim}
\email{jongkukkim@cau.ac.kr}
\affiliation{Department of Physics, Chung-Ang University, Seoul 06974, Korea.}
\affiliation{Excellence Cluster ORIGINS, Boltzmannstr. 2, D-85748 Garching, Germany.}
\author{Hyun Min Lee}
\email{hminlee@cau.ac.kr}
\affiliation{Department of Physics, Chung-Ang University, Seoul 06974, Korea.}

\begin{abstract} 
We propose a new model to explain the KM3NeT neutrino event through a low reheating scenario with a suppression in the GW spectrum originating from cosmic string networks. To achieve this, we extend the SM gauge sector by an abelian gauge symmetry and a singlet scalar. Once the abelian gauge symmetry spontaneously breaks, the extra gauge boson acquires mass and becomes a suitable Dark Matter (DM) candidate. Due to the kinetic mixing with the hypercharge gauge group, DM can decay into SM particles.
To explain the KM3NeT signal, we need $\mathcal{O}(100)$ PeV DM, which can be produced in the correct order of DM density in a low reheating scenario. In this scenario, the overabundance issue of heavy DM can be tackled by diluting its abundance through the continuous injection of entropy when the matter-like inflaton decays into the SM bath. Using the low reheating scenario, we can obtain the correct value of DM density both for freeze-out and freeze-in mechanisms for super-heavy DM.
Moreover, we have studied the Gravitational Waves (GWs) produced from cosmic strings, which fall within the detectable range of future proposed GW experiments. Additionally, the dominance of a quadratic inflaton potential before the reheating temperature changes the temperature–scale factor relation, which suppresses the GW spectrum at higher frequencies. Choosing an arbitrarily low reheating temperature provides only a tiny fraction of the DM density due to dilution from entropy injection.
This fraction of the vector DM suggests that only the extragalactic contribution is relevant in the KM3NeT event because DM lifetime is shorter than the age of the Universe. 
\end{abstract}
\maketitle

\section{Introduction}

With the progress of time, we have confirmed the validity of the Standard Model (SM) of particle physics with more and more precision. As it is evident, the SM cannot accommodate many aspects of beyond Standard Model (BSM) physics, important ones are the presence of dark matter and neutrino mass. So far, we have explored many BSM scenarios at different experiments, generally for the TeV scale DM candidate. Till date, we are unable to confirm the particle nature of DM and the origin of the neutrino mass. In this work, we will mainly focus on dark matter, but the neutrino mass can also be accommodated easily \cite{Costa:2022lpy}. There has been a rigorous search for DM in direct detection, indirect detection, and collider experiments, but the allowed regions have been shrinking without any detection. It is high time to look for the DM signal in other aspects, for example, using multimessenger astronomy.

In the context of multimessenger astronomy, the primary messengers include gravitational waves (GWs) generated by diverse astrophysical phenomena such as binary black hole coalescences, neutron star–black hole mergers, cosmic strings, and first-order phase transitions; neutrinos emitted from core-collapse supernovae, active galactic nuclei (AGNs), gamma-ray bursts (GRBs), and potentially dark matter decay or annihilation; and high-energy cosmic rays originating from AGNs, GRBs, and pulsars.
The important thing is that we can look for new physics in all of these searches. For detecting the multimessenger signals, we have state of the art detectors. For GW detection, we have ongoing and proposed experiments like SKA \cite{Janssen:2014dka}, EPTA \cite{EPTA:2011kjn}, LISA \cite{LISA:2017pwj, LISA:2022yao}, DECIGO/BBO \cite{Yagi:2011wg}, ET \cite{Punturo:2010zz}, CE \cite{Hild:2010id}, LIGO \cite{LIGOScientific:2014qfs, LIGOScientific:2016wof, LIGOScientific:2017adf}. For neutrinos, we have Super-Kamiokande \cite{Super-Kamiokande:1998qwk}, IceCube \cite{IceCube:2018fhm, IceCube:2020wum}, and KM3NeT \cite{KM3Net:2016zxf}. Finally, for high-energy photons, we have LHAASO-KM2A \cite{LHAASO:2023gne}, Extensive Air Shower-Moscow State University (EAS-MSU) \cite{Fomin:2017ypo}, and the Pierre Auger Observatory (PAO) \cite{Castellina:2019huz}.

In the present work, we look for the multimessenger signals in neutrinos and GWs, which have connections with DM. In the case of detecting high-energy neutrinos, we have well designed experiments like IceCube \cite{IceCube:2018fhm, IceCube:2020wum}, which has already detected very high-energy neutrinos up to the $\mathcal{O}(1)$ PeV. Recently, the KM3NeT collaboration \cite{KM3Net:2016zxf} has detected even more energetic neutrinos beyond what any other neutrino detectors have ever observed. KM3NeT has detected $\mathcal{O}(100)$ PeV neutrinos, which they labelled as the KM3-230213A signal, and around the same time IceCube, with an effective area approximately an order of magnitude larger than KM3-NeT, did not detect any high-energy neutrino like this. This is also one of the issues to be tackled. The only difference between the two detectors is the direction of the neutrino, which suggests that the neutrino/BSM particles have passed through 147 km of rock and sea, whereas for IceCube, it is only 14 km of ice.
Using this difference in location, there have been a few solutions in the literature to resolve the tension, such as sterile neutrinos that convert to active neutrinos when passing through the rock, as studied in Ref. \cite{Brdar:2025azm}, or up-scattering of BSM particles that eventually decay to $\mathcal{O}(70~{\rm PeV})$ muons \cite{Farzan:2025ydi, Dev:2025czz}. In the present work, we have considered the decaying DM scenario, which can predict the flux lower than the IceCube limit by suitably choosing the DM mass, lifetime, and its fraction compared to the total DM density \cite{Kohri:2025bsn}. In Ref. \cite{Kohri:2025bsn}, it has been shown that by considering super-heavy DM (SHDM), the tension between the KM3NeT signal and the non-observation at IceCube can be reduced to $1.2~\sigma$. In Ref. \cite{Kohri:2025bsn}, the authors pointed out that for DM mass in the range $M_{DM} = (1.5-52.0)\times 10^{8}$ GeV with lifetime $\tau_{DM} = (5.4-14.2)\times 10^{29}$ sec, one can explain the KM3NeT signal when DM directly decays dominantly into active neutrinos.
The present work uses SHDM which can directly decay into active neutrinos along with other SM particles. Moreover, there have been many attempts to explain the KM3NeT signal by considering different sources like extragalactic DM, PBHs, and other astrophysical sources, which can be found in Refs. \cite{Li:2025tqf, KM3NeT:2025vut, KM3NeT:2025bxl, KM3NeT:2025aps, KM3NeT:2025ccp, Boccia:2025hpm, Borah:2025igh, Brdar:2025azm, Kohri:2025bsn, Alves:2025xul, Crnogorcevic:2025vou, Klipfel:2025jql, Jho:2025gaf, Choi:2025hqt, Barman:2025hoz, Murase:2025uwv, Khan:2025gxs, He:2025bex, Baker:2025cff, Farzan:2025ydi, Dev:2025czz, Anchordoqui:2025xug, Airoldi:2025opo, Bertolez-Martinez:2025trs, Mondol:2025uuw, Yuan:2025zwe, Sakharov:2025oev,Palmisano:2025abd, Cermenati:2025ogl, Su:2025qzt, Aloisio:2025nts}.

In the present work, we have considered a SHDM scenario where we extend the SM sector by a dark abelian gauge symmetry $U(1)_D$ and a SM singlet scalar $\phi_D$ \cite{Lebedev:2011iq, Ko:2014gha, Hambye:2008bq, Khan:2024biq, Khan:2025keb, Khan:2025yoa}, which is charged under the same abelian gauge symmetry. The same scenario has been studied in the context of $\mathcal{O}(100)$ GeV scale DM for explaining different BSM signals like 511 keV gamma-ray line \cite{Khan:2024biq}. The gauge boson associated with the $U(1)_D$ can be a good DM candidate, whose longitudinal component is the CP-odd part of the extra singlet scalar. In general, by obeying the unitarity bound on the dark gauge coupling, we cannot accommodate DM mass beyond $M_{W_D} > 3\times 10^{5}$ GeV for the correct value of DM density, whereas in the freeze-in mechanism, such high mass can be accommodated but requires a very tiny gauge coupling. Moreover, we consider the kinetic mixing between $U(1)_{D}$ and the hypercharge gauge group, which is very important in our work for making our DM decay in order to explain the KM3NeT signal.

In this work, we consider a low reheating scenario of DM production \cite{Giudice:2000ex, Fornengo:2002db, Pallis:2004yy, Gelmini:2006pw, Drees:2006vh, Roszkowski:2014lga, Bernal:2018kcw, Arias:2021rer, Bernal:2022wck, Bhattiprolu:2022sdd, Haque:2023yra, Ghosh:2023tyz, Silva-Malpartida:2023yks, Bernal:2024yhu, Silva-Malpartida:2024emu, Barman:2024tjt,Bernal:2024ndy, Belanger:2024yoj}, where we assume the inflaton with a quadratic potential continuously decays to the SM bath, and subsequent interactions of the SM bath particles produce the DM candidate. During the reheating process, the inflaton continuously increases the SM entropy, which dilutes the overabundance of DM. Therefore, as will be discussed in the results section, we can accommodate $\mathcal{O}(100)$ PeV scale DM produced by the freeze-out mechanism, by obeying the unitarity bound on the gauge coupling \cite{Griest:1989wd}, as well as by the freeze-in mechanism for $\mathcal{O}(0.1)$ gauge coupling values. Irrespective of DM production through the freeze-out or freeze-in mechanism, DM can decay into the SM particle content through kinetic mixing and can explain the KM3NeT signal. In our scenario, the dominant decay modes of DM include SM quarks ($q\bar q$), leptons ($l\bar l$), and neutrinos ($\nu_l \bar \nu_l$).
With these choices of final states, and taking into account the branching of each channel,\footnote{DM decay to $W^{+}W^{-}$ is suppressed by the mutual cancellation of the mixing angles, which makes its branching subdominant.} we can easily explain the KM3NeT signal without producing flux beyond the IceCube limit. Additionally, due to the presence of SM quarks, our model is capable of explaining part of the IceCube events for suitable values of the DM mass and lifetime. We have used HDMSpectra \cite{Bauer:2020jay} for producing the neutrino and photon spectra from the decaying particles.

In discussing the phenomenology of SHDM, we study in detail the DM production and the response of DM relic density to variations in the model parameters. 
Moreover, we also show the allowed region after satisfying the DM constraints in different model parameter spaces. Finally, since we have $U(1)_{D}$ symmetry breaking with a very large vev due to SHDM, cosmic strings can be produced. The large vev increases the string tension, and we have found a detectable GW signal at future proposed GW experiments as discussed earlier. Moreover, as we consider a quadratic inflaton potential during reheating, the scale factor varies with the temperature as $a \propto T^{-3/8}$ \cite{Garcia:2020wiy}, which impacts GW spectrum by suppressing the GW relic at the higher frequencies. Therefore, if we observe such behaviour in the GW spectrum in the future, we can indirectly confirm the existence of such a reheating period in the early Universe.

The rest of the paper is structured as follows. In section \ref{sec:model}, we describe our DM model in detail. In section \ref{sec:thermalDM}, we discuss the overabundance problem of thermal DM for masses beyond 100 TeV. In section \ref{LRH-WIMP-FIMP}, we discuss the effect of low reheating temperature on DM abundance due to entropy injection. In section \ref{KM3NeT-signal}, we address the KM3NeT signal, and in section \ref{sec:GW} we show the detection prospects of the present model in future GW detectors. Finally, in section \ref{conclusion}, we conclude our work.

\section{Model}
\label{sec:model}
In this work, we consider a simple $U(1)_{D}$ extension of the SM gauge groups, along with an additional SM singlet scalar charged under the $U(1)_D$ gauge group. The complete particle content and gauge groups are presented in Table \ref{tab1}. In this framework, the $U(1)_{D}$ gauge boson is taken as the vector DM candidate, which acquires mass once the extra singlet scalar spontaneously develops a vev.

\begin{center}
\begin{table}[h!]
\begin{tabular}{||c|c|c|c||}
\hline
\hline
\begin{tabular}{c}
    Gauge\\
    Group\\ 
    \hline
    
    ${\rm SU(2)}_{\rm L}$\\  
    \hline
    ${\rm U(1)}_{\rm Y}$\\
    \hline
    ${\rm U(1)}_{\rm D}$\\      
\end{tabular}
&

\begin{tabular}{c|c|c}
    \multicolumn{3}{c}{Baryon Fields}\\ 
    \hline
    $Q_{L}^{i}=(u_{L}^{i},d_{L}^{i})^{T}$&$u_{R}^{i}$&$d_{R}^{i}$\\ 
    \hline
    $2$&$1$&$1$\\ 
    \hline
    $1/6$&$2/3$&$-1/3$\\
    \hline
    $0$&$0$&$0$\\      
\end{tabular}
&
\begin{tabular}{c|c}
    \multicolumn{2}{c}{Lepton Fields}\\
    \hline
    $L_{L}^{i}=(\nu_{L}^{i},e_{L}^{i})^{T}$ & $e_{R}^{i}$\\
    \hline
    $2$&$1$\\
    \hline
    $-1/2$&$-1$\\
    \hline
    $0$&$0$\\    
\end{tabular}
&
\begin{tabular}{c|c}
    \multicolumn{2}{c}{Scalar Field}\\
    \hline
    $H$ & $\phi_{D}$\\
    \hline
    $2$ & $1$\\
    \hline
    $1/2$ & $0$\\
    \hline
    $0$ & $1$\\    
\end{tabular}\\
\hline
\hline
\end{tabular}
\caption{SM particle contents and extra BSM scalar and their corresponding
charges under SM and $U(1)_D$ gauge groups. }
\label{tab1}
\end{table}
\end{center}

The complete Lagrangian consisting of SM and BSM particles is given by
\begin{align}
    \mathcal{L} &= 
    \mathcal{L}_{{\rm SM}}
    +\biggl(D^{\dagger}_{\mu}\phi_D D^{\mu} \phi_D \biggr)
    - \mathcal{L}^{gauge}_{Kinetic}
    -\mathcal{V}(\phi_D,H),
\end{align}
where $\mathcal{L}_{\rm SM}$ is the Lagrangian consisting of the SM particles, the next term is the kinetic term for the singlet scalar, and $\mathcal{L}^{gauge}_{Kinetic}$ is the kinetic term involving $U(1)_D$ and $U(1)_{Y}$ as well as their mixing, which is given in Eq. \ref{kinetic-terms} and will be discussed in detail in section \ref{KM3NeT-signal}. Finally, the last term, $\mathcal{V}(\phi_D,H)$, is the potential involving the singlet scalar and the SM Higgs doublet, which takes the following form,
\begin{align}
    V(\phi_D,H) = 
    - \mu^2_D \phi^{\dagger}_D \phi_D
    + \lambda_D (\phi^{\dagger}_D \phi_D)^2 
    - \mu^2_H H^{\dagger} H
    + \lambda_H (H^{\dagger} H)^2 
    + \lambda_{HD} \phi^{\dagger}_D \phi_D H^{\dagger} H
    \,.\label{eqn:scalar_pot_full}
\end{align}
During the Universe evolution, the scalars acquire spontaneous vevs which take the following form in the Unitary gauge,
\begin{align}
    H = \begin{pmatrix}
        0\\
        \frac{v_{h} + h}{\sqrt{2}}
    \end{pmatrix},~~
    \phi_D = \frac{v_{D} + \phi_0}{\sqrt{2}}.
\end{align}
Once the scalars acquire vevs, the scalar mass matrix in the basis ($h ~~\phi_0$) can be expressed as follows,
\begin{eqnarray}
    M_{h\phi_0} = \begin{pmatrix}
        2 \lambda_{H} v^2_{h} & \lambda_{HD} v_{D} v_{h}\\
        \lambda_{HD} v_{D} v_{h} & 2 \lambda_{D} v^2_{D} 
    \end{pmatrix}
\end{eqnarray}
We can diagonalise the above mass matrix and define mass basis $(h_{1}~~h_{2})$ in terms of the flavour basis $(h~~\phi_0)$ through the following equations,
\begin{align}
    \begin{pmatrix}
        h_{1}\\
        h_{2}
    \end{pmatrix}
    = \begin{pmatrix}
        \cos\theta & - \sin\theta \\
        \sin\theta & \cos\theta
    \end{pmatrix}
    \begin{pmatrix}
        h\\
        \phi_0
    \end{pmatrix}.
\end{align}
Moreover, the quartic couplings in the mass matrix can be expressed in terms of the masses and the mixing angle as follows,
\begin{align}
    \lambda_{H} &= \frac{M^2_{h_1} \cos^{2}\theta + M^2_{h_2} \sin^{2}\theta }{2 v^2_{h}}, \nonumber \\
    \lambda_{D} &= \frac{M^2_{h_1} \sin^{2}\theta + M^2_{h_2} \cos^{2}\theta }{2 v^2_{D}}, \nonumber \\
    \lambda_{HD} &= \frac{ \biggl( M^2_{h_2} - M^2_{h_1} \biggr) \sin\theta \cos\theta }{v_{h} v_{D}} 
\end{align}
Finally, the vector DM also acquires mass, whose longitudinal component comes from the CP-odd component of the singlet scalar, and takes the form $M_{W_D} = g_{D} v_{D}$.

\section{WIMP Dark Matter without low reheating}
\label{sec:thermalDM}
Before moving to the discussion on the low reheating framework, we briefly outline the limitations of producing heavy DM via the standard freeze-out mechanism. As discussed in Ref. \cite{Griest:1989wd}, due to the unitarity bound on the relevant parameters, it is difficult to obtain the correct DM relic density for $M_{W_D} > 10^{5}$ GeV. In our study of $U(1)_{D}$ vector DM, the dominant annihilation mode is $W_{D} W_{D} \rightarrow h_{2} h_{2}$, while other processes are subdominant.
\begin{figure}[h!]
\centering
\includegraphics[angle=0,height=8.5cm,width=10.5cm]{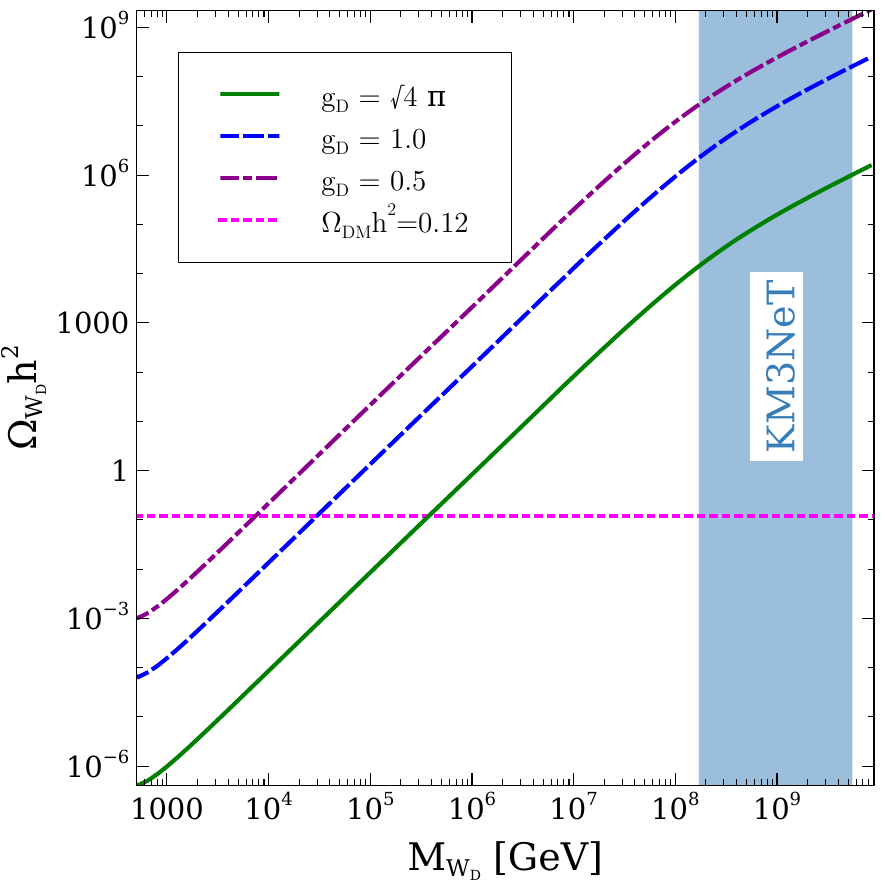}
\caption{Variation of DM relic density with the DM mass for three different values of the gauge coupling. The shaded violet region denotes the DM mass range that can explain the KM3NeT signal.
Dashed magenta line correspond to the DM relic density from Planck data, $\Omega_{\rm DM}h^2=0.12$.
}
\label{thermal-DM-RD-variation}
\end{figure}

The Boltzmann equation relevant for estimating the DM relic density using the freeze-out mechanism is,
\begin{align}
\frac{d Y_{W_D}}{d x} = - \frac{s(x) \langle \sigma v \rangle}{x H(x)}
\left[ Y^{2}_{W_D} - \left( Y^{\rm eq}_{W_D} \right)^2 \right],\qquad
\label{BE}
\end{align}
where $x=\frac{M_{W_D}}{T}$. The thermal average of the dominant annihilation mode $\langle \sigma v \rangle$ is given in Eq. \ref{sigvWDWD}, and the entropy density $s(x)$ and the Hubble parameter $H(x)$ are expressed as \cite{Gondolo:1990dk,Edsjo:1997bg},
\begin{align}
s(x) = \frac{2 \pi^2}{45} g_{s} M^3_{W_D} x^{-3},\quad
H(x) = \sqrt{\frac{g_ \pi^2}{90}} \frac{M^2_{W_D}}{M_{\rm Pl}} x^{-2}.
\end{align}
To solve the above equation, it is necessary to estimate the freeze-out temperature $x_f$, obtained from the condition,
\begin{align}
n^{\rm eq}_{W_D} \langle \sigma v \rangle \simeq H(x)~~ {\rm at}~~x=x_f. 
\label{equilibrium-condition}
\end{align}
Once the freeze-out temperature $x_f$ is determined, the Boltzmann equation in Eq. \ref{BE} can be solved analytically to estimate the co-moving number density, which takes the following form,
\begin{align}
Y^0_{W_D} = Y^{f}_{W_D} \left[
1 + \frac{2\pi^2}{45} \sqrt{\frac{90}{\pi^2}} M_{\rm P} m_i Y^f_{W_D}
\frac{g_{s}(x^f)}{\sqrt{g_(x^f)}} \frac{\langle \sigma v \rangle_{W_{D}W_{D}\rightarrow h_{2}h_{2}}}{x_f}
\right]^{-1},
\label{eqn:YieldToday}
\end{align}
where $\langle \sigma v \rangle_{W_{D}W_{D}\rightarrow h_{2}h_{2}}$ is defined in Eq. \ref{sigvWDWD}, and the equilibrium co-moving yield $Y^{0}_{W_D}$ today ($t_{0}$) is given by
\begin{align}
Y^f_{W_D} = \frac{45}{2\pi^4} \sqrt{\frac{\pi}{8}} \frac{g_i}{g_{s}(x_f)}(x_f)^{3/2}e^{-x_f}.
\end{align}
Here, $Y^{f}_{W_D}$ at $x=x_{f}$ denotes the equilibrium DM density at freeze-out.
Finally, the DM relic density can be obtained from the co-moving number density using \cite{Edsjo:1997bg}
\begin{align}
\Omega_{W_D} h^2= 2.755\times 10^{8}~ Y^{0}_{W_{D}}~
\Big(
\frac{M_{W_D}}{\rm GeV}
\Big).
\end{align}

In Fig. \ref{thermal-DM-RD-variation}, we show the variation of the DM relic density with the DM mass. From Eq. \ref{ab-psi-wd}, we can see the thermally averaged annihilation cross section decreases as the DM mass increases, leading to an enhancement in the relic density, consistent with the figure. For the maximum gauge coupling allowed by unitarity, $g_{D} = \sqrt{4 \pi}$, DM is overproduced for $M_{W_D} > 3 \times 10^{5}$ GeV. For smaller values of $g_D$, overproduction occurs at lower masses, for example, for the gauge coupling value $g_{D} = 0.1$, we obtain $\Omega_{W_D}h^{2} > 0.12$ for $M_{W_D} > 10^{4}$ GeV.
It follows that it is not possible to accommodate very heavy DM in the thermal freeze-out scenario beyond $M_{W_D} > 3\times 10^{5}$ GeV. 
In the plot, we also highlight the decaying DM mass range capable of explaining the recent KM3NeT signal \cite{KM3NeT:2025npi} if DM decays directly to neutrinos along with other SM and BSM states. Clearly, the standard thermal scenario cannot simultaneously explain both the KM3NeT signal and the observed relic abundance for heavy DM. 
In this work, we propose a novel mechanism to overcome this limitation by introducing entropy injection from inflaton decay, which continuously dilutes the DM density until the temperature drops to the reheating temperature, thereby yielding the correct relic abundance.
\section{WIMP/FIMP Dark Matter with low reheating}
\label{LRH-WIMP-FIMP}
As discussed earlier, it is difficult to obtain the correct relic density for $\mathcal{O}({\rm PeV})$-scale DM in the standard cosmological scenario. In this section, we consider DM production in the non-standard cosmology scenario, where we assume that after the end of inflation, the inflaton oscillates and produces the SM bath, and subsequently DM is produced from the radiation bath.
In this context, we consider a quadratic inflation potential, which can be derived from a Starobinsky-type potential \cite{Starobinsky:1980te, Khan:2025kuh} once the inflaton field value becomes sub-Planckian during reheating. Furthermore, because our analysis is restricted to quadratic inflation, we do not include any effects of inflaton fragmentation in our study, 
which arise for polynomial inflaton potentials with $k\geq 4$ where
$k$ denotes the power of the inflaton field in the inflation potential. Between the end of inflation and the beginning of reheating, there can be a phase called preheating, where perturbative calculations do not hold. In the present work, we have not considered the possibility of violent production of DM during the preheating era. The preheating era and DM production depend on the full inflation set-up and on how the inflaton is related to the DM. In general, we can keep the inflaton and DM independent or choose parameters such that the inflaton decay to DM is not kinematically allowed during the preheating era \cite{Khan:2025kuh}. Therefore, to incorporate preheating effects, one needs a complete inflation set-up, which is not required in the present work, hence we can formulate a set-up where the inflaton and DM are not directly connected. In Ref.~\cite{Ema:2016dny}, it has been shown that if the DM longitudinal component comes from the CP-odd part of the inflaton, then DM will undergo violent production during the preheating era.
Moreover, to realise the preheating era, one needs particular set of parameters value of the inflation potential, as shown in Ref.~\cite{Ema:2016dny}. Therefore, connecting the full inflationary dynamics to the present kind of DM study requires a dedicated study and will be explored in the future. In this work, we only focus on the reheating era, where the inflaton gradually decays and a radiation bath is produced, which subsequently produces DM.
This process of producing the SM bath can happen at a late time, making the low reheating temperature feasible. We assume that during the decay of the inflaton $\phi$, it behaves like a matter field {\it i.e.} the potential of the inflaton field is,
\begin{align}
    V(\phi) = \lambda_{\phi} M^2_{pl} \phi^2.
\end{align}
The coupled differential equations for the inflaton energy density and the SM entropy density can be expressed as follows \cite{Gelmini:2006pw},
\begin{align}
\frac{d \rho_{\phi}}{dt} + 3 H \rho_{\phi} &= - \Gamma_{\phi} \rho_{\phi}, \nonumber \\
\frac{ds}{dt} + 3 H s &= \frac{\Gamma_{\phi} \rho_{\phi} }{T}
\label{BE-inf-ent}
\end{align}
where $\rho_\phi$ is the inflaton energy density, and $s(T) = \frac{2 \pi^2}{45} h_{eff}(T) T^{3}$ is the entropy density, with $h_{eff}(T)$ being the entropic relativistic d.o.f. Once we have the temperature information from the entropy, we can estimate the total energy density of the SM radiation bath using the following expression,
\begin{align}
\rho_{R} = \frac{\pi^2}{30} g_{eff}(T) T^{4},
\end{align}
where $g_{eff}(T)$ is the relativistic {\it d.o.f} for the matter part.
It is clear from Eq. \ref{BE-inf-ent} that the inflaton energy density decreases with the Universe's expansion, while the SM entropy/radiation density increases. Moreover, the inflaton and radiation energy densities are related through the Friedmann equation as follows,
\begin{align}
H^{2} = \frac{8 \pi}{3} \frac{\rho_{\phi} + \rho_{ R}}{M^2_{pl}}
\end{align}
where $M_{pl} = 1.22\times 10^{19}$ GeV is the Planck mass. As discussed in Ref. \cite{Belanger:2024yoj}, it is convenient to define new parameters as $z_{\phi} = \rho_{\phi} a^{3}$ and $z_{s} = s^{4/3} a^{4}$, where $a$ is the scale factor related to the Hubble parameter as $H = \frac{\Dot{a}}{a}$. This parametrisation reduces the above set of differential equations in Eq. \ref{BE-inf-ent} to,
\begin{align}
\frac{d z_{\phi}}{da} &= -\frac{\Gamma_{\phi}}{H a} z_{\phi}, \nonumber \\
\frac{d z_{s}}{da} &= \frac{4}{3} \frac{\Gamma_{\phi}}{H} \frac{s^{1/3}}{T} z_{\phi}.
\end{align}
The above differential equations can be solved with initial conditions, such as at $t=t_{I}$, scale factor $a_{I}=a(t_{I})=1$, $\rho_{R}(t_{I})=0$, and $\rho_{\phi}(t_{I}) = \frac{3}{8\pi} M^2_{pl} H^2_{I}$, where $H_{I}$ is the inflation scale, constrained by the non-observation of B-mode polarisation to $H_{I} < 4 \times 10^{-6} M_{pl}$ \cite{BICEP:2021xfz}. The important parameter in our study is the inflaton decay width $\Gamma_{\phi}$, which determines the reheating temperature ($T_{\rm reh}$) through the perturbative reheating approximation ($\Gamma_{\phi} = H$) given by,
\begin{align}
T^2_{reh} = \frac{3}{2\pi} \sqrt{\frac{5}{\pi g_{eff}(T_{\rm reh})}} M_{pl} \Gamma_{\phi}
\end{align}
Moreover, the temperature of the produced radiation bath varies with the scale factor and attains a maximum temperature as follows \cite{Barman:2021ugy},
\begin{align}
T^4_{max} = \frac{15}{2 \pi^{3} g_{eff}(T_{max})} \left( \frac{3}{8} \right)^{8/5} M^2_{pl}
\Gamma_{\phi} H_{I}
\label{Tmax}
\end{align}
In our study, we choose $H_I$ such that $T_{\text{max}}$ lies above all particle masses considered. It should be emphasised that the two reference temperatures, $T_{\text{reh}}$ and $T_{\text{max}}$, play a crucial role in our analysis. 
In Refs.~\cite{Clery:2021bwz, Haque:2021mab}, it has been pointed out that for a reheat temperature $T_{reh} \leq 10^{9}$ GeV, the 
gravitational-portal production of the radiation bath from inflaton annihilation dominates over direct inflaton decay for a short period of time at the very beginning of reheating. This dominant gravitational production at early times is unavoidable and sets the maximum temperature of the Universe, which takes the following form for the quadratic inflaton potential, 
\begin{eqnarray}
    T^{grav}_{max} \simeq 3 \times 10^{12} \left( \frac{\rho_{end}}{10^{64}~{\rm GeV}^{4}} \right)^{3/8}
    \left( \frac{m_{\phi}}{3 \times 10^{13}~{\rm GeV}} \right)^{1/4}~{\rm GeV}.
    \label{Tmax-grav}
\end{eqnarray}
If $T_{reh} < 10^{9}$ GeV, the true maximal temperature is larger than what we would estimate by Eq. \ref{Tmax}. But, as far as the maximal temperature is much larger than the freeze-out temperature, as in our case, our results are not affected.
It is to be noted that we have used micrOMEGAs for the production of DM, which uses Eq.~\ref{Tmax} in determining $T_{max}$, and we have ensured that the masses of the relevant particle spectrum in our set-up are always smaller than $T_{max}$.
Additionally, as is well known, for UV complete theory, freeze-in DM production is independent of the initial temperature \cite{Biswas:2016yjr, Biswas:2017tce} as long as it is above the relevant mass scale, therefore, the DM abundance remains unchanged. Hence, our results do not change even if we consider $T_{max} \sim 10^{12}$ GeV by following Eq. \ref{Tmax-grav}, because the DM masses have been chosen to satisfy the condition $T_{max} > M_{W_D}$. We have verified that the particle masses in our spectrum always remain below $T_{\text{max}}$, ensuring that no Boltzmann suppression occurs for any of them, since the number density is proportional to the modified Bessel function of the second kind, {\it i.e.} $n_{eq} \propto K_{2}\left(\frac{M_{W_D}}{T}\right)$, which exhibits suppression when $T < M_{W_D}$, while dominant DM production occurs when $T \sim M_{W_D}$. In the case of WIMP-type DM, we have focused on the DM coupling to the SM sector in such a range that we do not have ultra-relativistic freeze-out (UFO), which happens at temperature $T \geq M_{W_D}$ \cite{Henrich:2025gsd}. In the UFO scenario, which is not applicable in the present work, $T_{max}$ might affect the DM relic density because in some cases the relativistic freeze-out temperature ($T^{U}_{FO}$) can be larger than the maximum temperature, {\it i.e.} $T^{U}_{FO} > T_{max} > M_{W_D}$, which will modify the DM relic density estimation due to the unavailability of such high temperature.
In Refs.~\cite{Cosme:2023xpa, Cosme:2024ndc, Arcadi:2024wwg, Boddy:2024vgt, Lee:2024wes,Mondal:2025awq, Khan:2025keb}, authors have used this Boltzmann suppression property to study freeze-in DM at stronger couplings. 
In the present work, we consider DM density dilution through entropy production until the reheating temperature $T_{\rm reh}$, 
after reheating, the DM relic abundance gets frozen at a particular value. We have used the built-in micrOMEGAs (v6.2.3) \cite{Alguero:2023zol} routines to estimate DM abundances in the low reheating scenario as discussed in detail in Ref. \cite{Belanger:2024yoj}.
It is to be noted that to check the thermality of DM, in Ref. \cite{Belanger:2024yoj}, the authors considered that if $|\delta z_{W_D}| < 10^{-2} \bar z_{W_D}$ and $\delta a < 10^{-3} a$, then DM is thermal; otherwise, it is non-thermal which is used in micrOMEGAs and we have followed it.
In the aforementioned constraints, the variations of $z_{DM}$ and $a$ can be expressed as \cite{Belanger:2024yoj},
\begin{align}
\delta z_{W_D} = -\frac{d \log(\bar z_{W_D})}{da} \frac{H a^{4}}{2 \langle \sigma v \rangle},~~
\delta a = \frac{1}{\bar z_{W_D}}\frac{H a^{4}}{ 2 \langle \sigma v \rangle}.
\label{thermality-check}
\end{align}
As defined earlier, $\bar z_{W_D} = n^{eq}_{W_D} a^{3}$, where $\langle \sigma v \rangle$ denotes the thermal averaged DM annihilation cross section. Depending on the Hubble parameter $H$, which in turn depends on $\Gamma{_\phi}$, DM can be either thermal or non-thermal even for the same interaction strength $\langle \sigma v \rangle$.
Finally, based on the $|\delta z_{W_D}|$ and $\delta a$ criteria, the Boltzmann equations will be solved in micrOMEGAs with either the equilibrium abundance or zero abundance as the initial condition at $t = t_{I}$.
It is worth mentioning that in the present work we have not considered DM production from inflaton annihilation mediated by the graviton, which is unavoidable. Following Refs.~\cite{Cook:2015vqa, Clery:2021bwz, Haque:2021mab, Haque:2020zco, Lee:2024rjw}, we have checked that DM produced through the gravitational portal from inflaton annihilation is suppressed compared to the usual freeze-in production from thermal bath particles. We have verified this for Starobinsky-type potential which can be approximated to the quadratic inflaton potential in the limit of sub-Plackian field value during reheating, and found that the gravitationally produced DM is always subdominant; therefore, we have not included gravitational production in the present work. It is to be noted that this conclusion holds for the quadratic inflaton potential and for the inflation potential with a higher power of the inflaton field may dominantly produce the DM through the gravitational portal as discussed in  Refs. \cite{Clery:2021bwz, Haque:2021mab, Haque:2020zco}. The gravitational production of DM is relevant only when DM is produced via the freeze-in mechanism. Once DM is produced through the freeze-out mechanism, we consider the DM number density to be same as the thermal density, and the prior history of its production is not relevant for the WIMP-type DM candidate.

As we discussed in section \ref{sec:thermalDM}, in the perturbative regime with gauge coupling value $g_{D}<\sqrt{4 \pi}$, DM is always overproduced for $M_{W_D} > 3 \times 10^{5}$ GeV. In the following plots, we discuss how it is possible to accommodate heavy DM candidates in the low reheating temperature scenario due to the continuous production of entropy, which will be suitable to explain the KM3NeT signal. 
One important aspect in DM studies is that DM must be cold during the structure formation era of the Universe, {\it i.e.} at temperatures $T \sim 1$ eV.
In Refs.~\cite{Viel:2013fqw, Narayanan:2000tp, Viel:2005qj, Baur:2015jsy, Irsic:2017ixq, Palanque-Delabrouille:2019iyz, Garzilli:2019qki}, the authors have placed bounds on the warm DM mass, which is mostly above a few keV and can be as low as $M_{W_D} > 1.9$ keV, below this value, DM will contradict the Lyman-$\alpha$ data because of the larger values of the free streaming length of DM.
In Refs.~\cite{Haque:2021mab, Henrich:2025gsd, Henrich:2025sli}, the authors have studied the effect of warm DM on structure formation in the low reheating temperature scenario. Additionally, in Ref.~\cite{Henrich:2025gsd}, authors have shown that for UFO DM, when the freeze-out temperature is larger than the reheating temperature, there will be additional suppression in the temperature of DM, which can accommodate even lower masses than the value mentioned earlier (Refs. \cite{Choi:2019osi, Eroncel:2025qlk} also discuss the lower masses satisfying the Lyman-$\alpha$ bound.).
In the present work, we focus on the $\mathcal{O}(\text{PeV})$-scale DM candidate to explain the KM3NeT signal, which is non-relativistic during its production and therefore easily evades the aforementioned bounds. Therefore, the parameter space considered in this work is safe from both the Lyman-$\alpha$ bound arising from structure formation and the negligible contribution to the relativistic degrees of freedom during BBN.
In the case of freeze-in DM, if DM has higher momentum than its mass during the production time, then it can impact structure formation; however, in our case, the abundance of heavy vector DM 
freezes out while DM is non-relativistic, so there is no problem for structure formation, unlike the scenario where a long-lived particle decays later into heavy DM and imparts momentum larger than the associated temperature of the Universe \cite{Decant:2021mhj, Covi:1999ty}.

\begin{figure}[h!]
\centering
\includegraphics[angle=0,height=7.5cm,width=8.5cm]{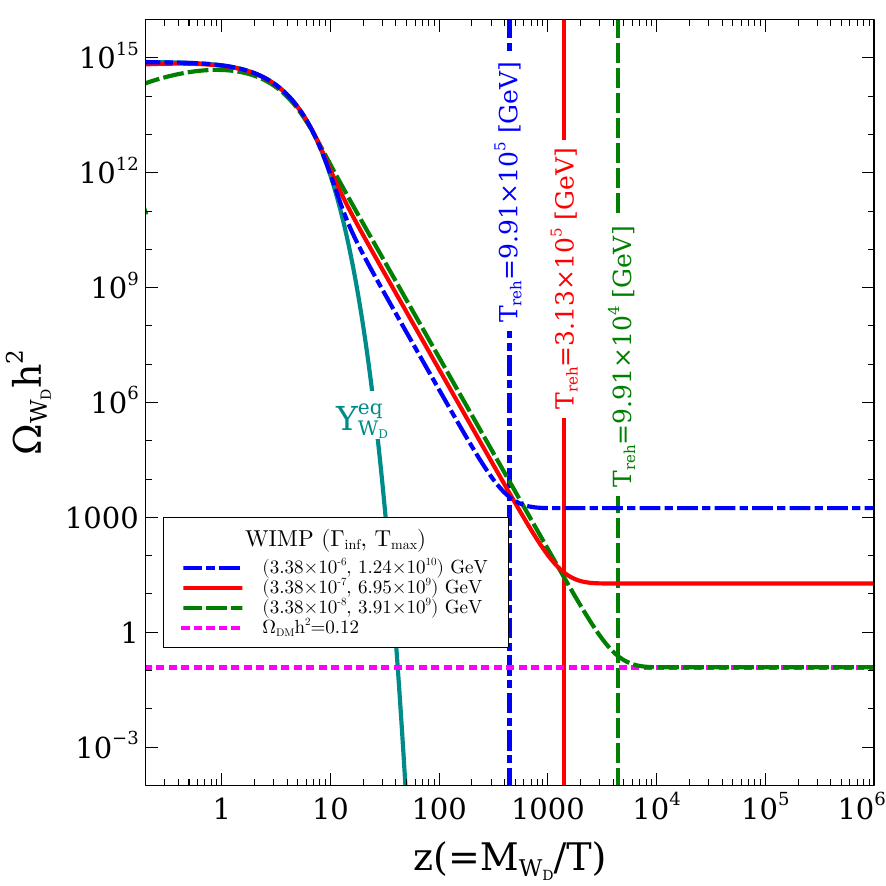}
\includegraphics[angle=0,height=7.5cm,width=8.5cm]{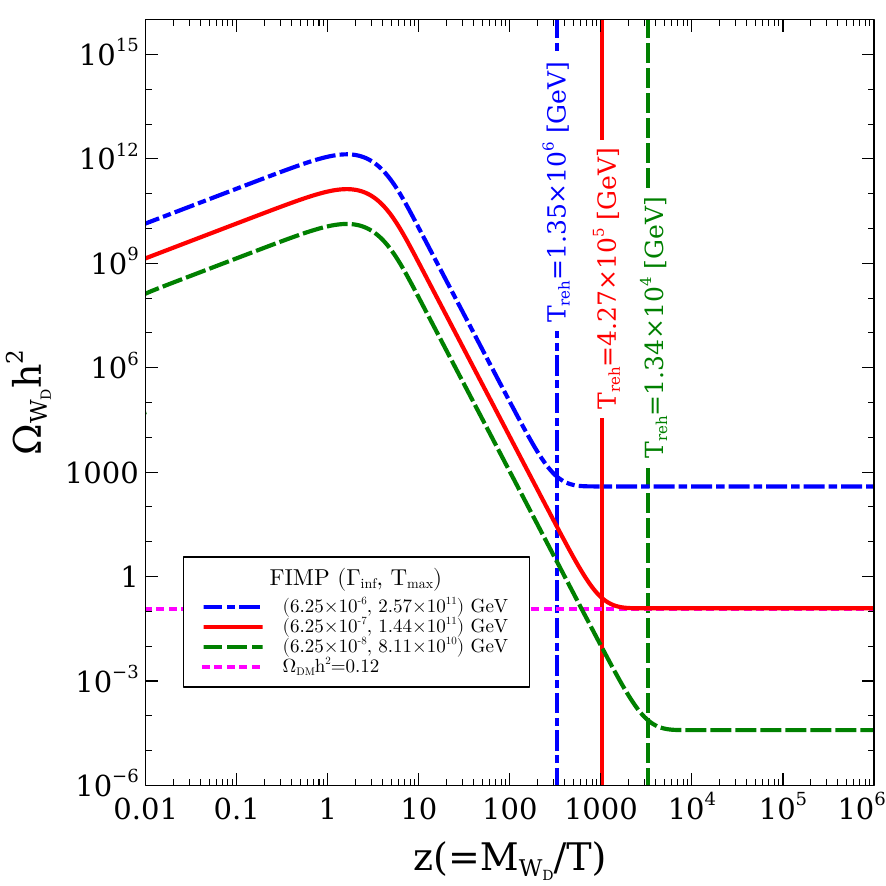}
\caption{LP shows DM production via the freeze-out mechanism, and RP shows DM production via the freeze-in mechanism in the low reheating scenario. Both plots are presented for three different values of the inflaton decay width $\Gamma_{\rm inf}$. The other parameter values have been kept fixed for LP (RP) at $\sin\alpha = 0.1$, $M_{h_2} = 300$ GeV, $g_{D} = 2.0$ ($g_{D}=0.1$), and $H_{I} = 10^{11}$ GeV.} 
\label{line-plot-FO-FI}
\end{figure}

In Fig.~\ref{line-plot-FO-FI}, we show DM production via the freeze-out and freeze-in mechanisms as a function of $x(=M_{W_D}/T)$ for three different values of $\Gamma_{\rm inf}$. In the LP, we show WIMP-type DM produced via the freeze-out mechanism. In the standard scenario, WIMP DM freezes out around $x=20$-$30$, which overproduces the DM density for $M_{W_D} > 3\times10^{5}$ GeV, as discussed earlier.
In this work, we consider a low reheating temperature scenario in which the SM bath is continuously produced through inflaton decay, resulting in entropy production and a dilution of the DM density. 
From the LP, we see that DM departs from thermal equilibrium around $z \simeq 20$ and is continuously diluted until the reheating temperature $T_{\text{reh}}$, totally determined by the inflaton decay width (blue dash–dot, red solid, and green dashed curves).
The plot clearly shows that lowering the reheating temperature (corresponding to a smaller inflaton decay width) leads to greater dilution of the DM relic density. This dilution allows the DM relic density to reach the correct order beyond the mass limit obtained from the unitarity bound.
In the plot, the magenta dotted line represents the correct DM density from Planck data, while the cyan line shows the DM density evolution if it remained in thermal equilibrium until today. For $z<20$, DM follows the equilibrium distribution, and around $z \sim 20$, it freezes out and is diluted due to entropy production until the reheating temperature. After $T_{\rm reh}$, the DM abundance freezes at a particular value. This figure implies that with a suitable choice of $T_{\rm reh}$ and gauge coupling $g_D$, we can achieve the correct DM density for $M_{W_D} > 10^{5}$ GeV.
In the RP, we show the DM relic density evolution for the freeze-in mechanism. From the plot, we see that DM is mainly produced around $z \sim 1$, corresponding to a temperature near the DM mass. DM continues to be produced from the thermal bath until $z \sim 1$, after which the relic density is diluted until the reheating temperature. By appropriately choosing $T_{\rm reh}$, we can achieve the correct DM relic density for heavy DM with a reasonable gauge coupling value.
The plot also shows that a higher value of $\Gamma_{\rm inf}$ leads to greater DM production for a given $z$, implying earlier radiation bath production and longer DM production until $T \sim M_{W_D}$. In the subsequent plots, we discuss how different model parameters affect the DM relic density.

\begin{figure}[h!]
\centering
\includegraphics[angle=0,height=7.5cm,width=8.5cm]{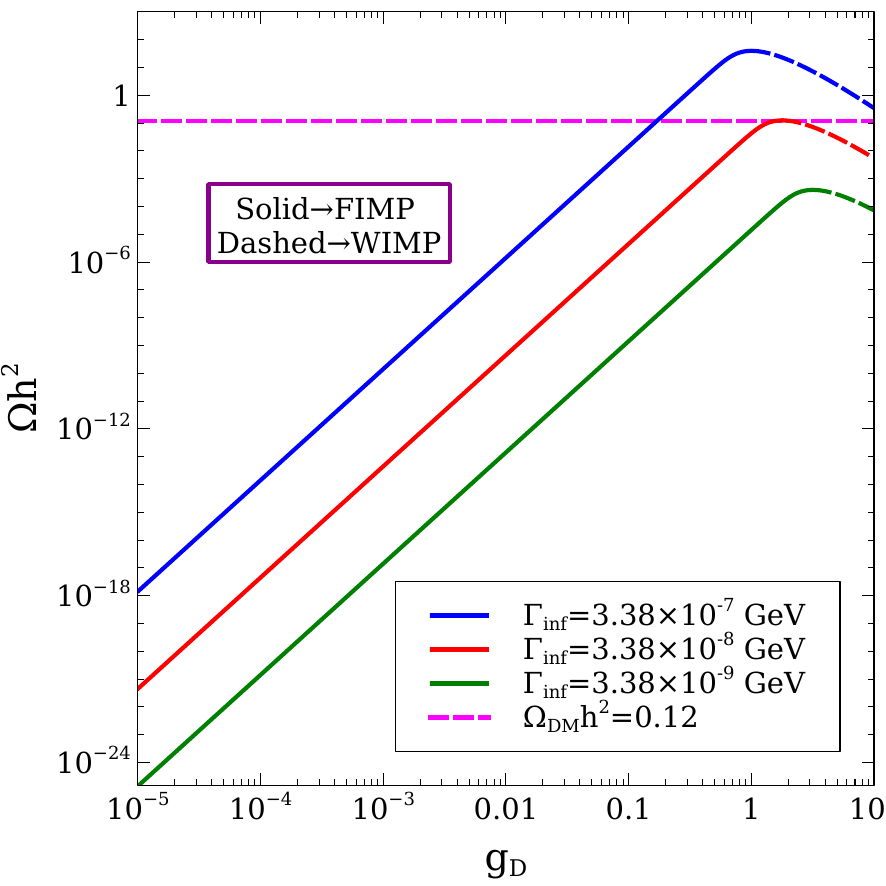}
\includegraphics[angle=0,height=7.5cm,width=8.5cm]{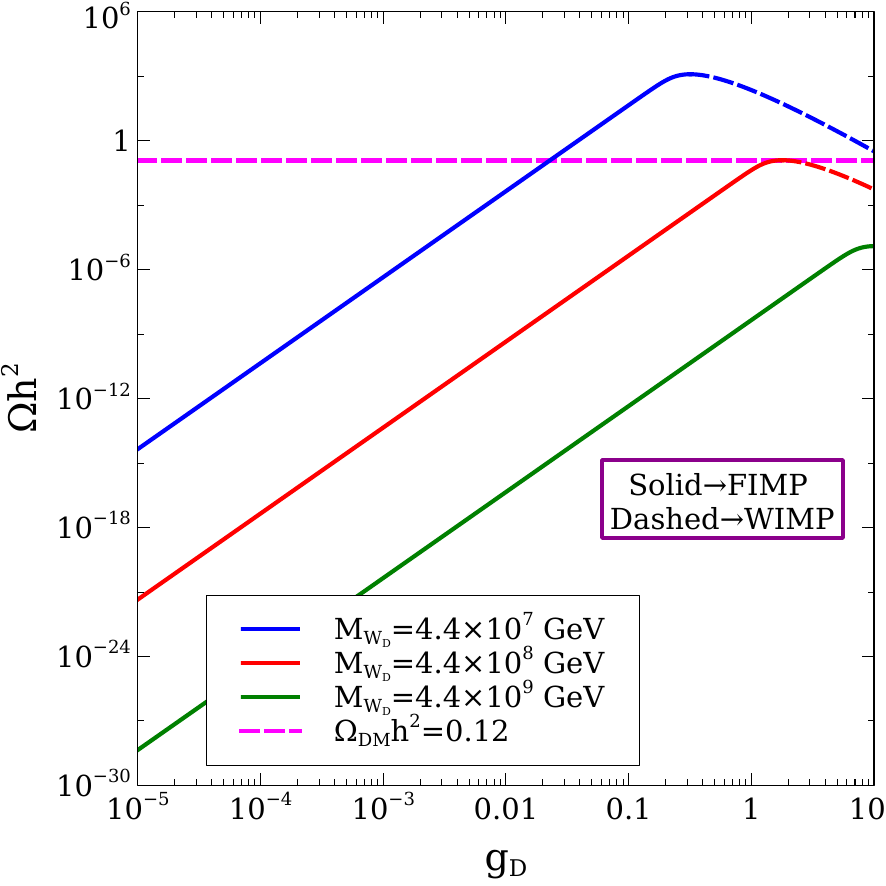}
\caption{LP and RP show the variation of DM relic density with the gauge coupling for three values of the inflaton decay width $\Gamma_{\rm inf}$ and DM mass $M_{W_D}$, respectively. The solid line shows DM production via the freeze-in mechanism, corresponding to FIMP-type DM, while the dashed line shows the freeze-out mechanism, corresponding to WIMP-type DM. The other parameter values have been kept fixed at $M_{h_2} = 300$ GeV, $M_{W_D} = 4.4 \times 10^{8}$ GeV, $g_{D} = 2$, $\Gamma_{\rm inf}=3.38 \times 10^{-8}$ GeV, $\sin\alpha=0.1$, and $H_{0} = 10^{11}$ GeV, unless the parameters are varied.} 
\label{line-plot-1}
\end{figure}

In Fig. \ref{line-plot-1}, we show the variation of DM relic density with the $U(1)_D$ gauge coupling $g_D$. The smooth transition of the DM production mechanism from freeze-in to freeze-out is described by the transition from solid to dashed lines. In the LP, we present the variation for three different values of the inflaton decay width, whereas in the RP, we show it for three different values of the DM mass. 
The caption contains the other parameter values, which remain fixed unless they are varied, and these are the same for the other line plots as well.
In the LP, we can see that the DM relic density increases with increasing gauge coupling for freeze-in DM (solid line), while for WIMP-type DM (dashed line), the relic density decreases as the gauge coupling increases. 
This relic density dependence on the gauge coupling is generic for freeze-in and freeze-out DM candidates, and the same conclusion holds for the RP.
In the LP, we illustrate the variation of the DM relic density for three different values of the inflaton decay width, $\Gamma_{\text{inf}}$. As shown in the plot, a smaller $\Gamma_{\text{inf}}$ corresponds to a lower reheating temperature, allowing more time for entropy production and thereby reducing the DM relic density.
Furthermore, the value of $\Gamma_{\text{inf}}$ has an effect on the transition point from FIMP to WIMP, which shifts to the right due to the longer inflaton decay time, influencing the Hubble parameter and, consequently, the thermality condition of WIMP DM.
In the RP, we show the relic density variation for three different values of DM mass $M_{W_D}$. In the freeze-out regime, as the DM mass increases, the relic density decreases because the DM relic density is proportional to the thermally averaged cross section $\langle \sigma v \rangle_{h_2 h_2 \rightarrow W_{D} W_{D}}$, which itself is proportional to $1/M^2_{W_D}$, as can be realised from  Eq. \ref{ab-psi-wd}. We can also notice that as $M_{W_D}$ increases, the FIMP-to-WIMP transition shifts to the right, compensating for the effect of the increased DM mass on the thermal average cross section by requiring a larger gauge coupling. Furthermore, for $M_{W_D} = 4.4 \times 10^{9}$ GeV, there is no DM production via the freeze-out mechanism, because at large $M_{W_D}$ values, $\langle \sigma v \rangle$ becomes sufficiently small, requiring gauge coupling values beyond the perturbative regime to maintain equilibrium.

\begin{figure}[h!]
\centering
\includegraphics[angle=0,height=7.5cm,width=8.5cm]{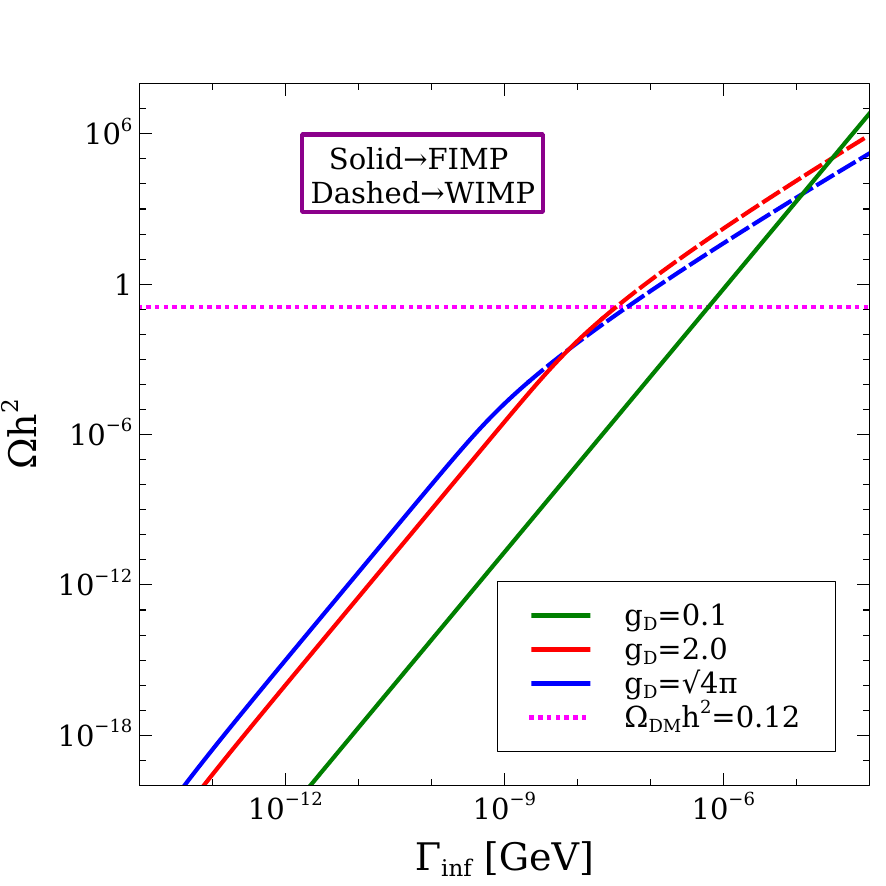}
\includegraphics[angle=0,height=7.5cm,width=8.5cm]{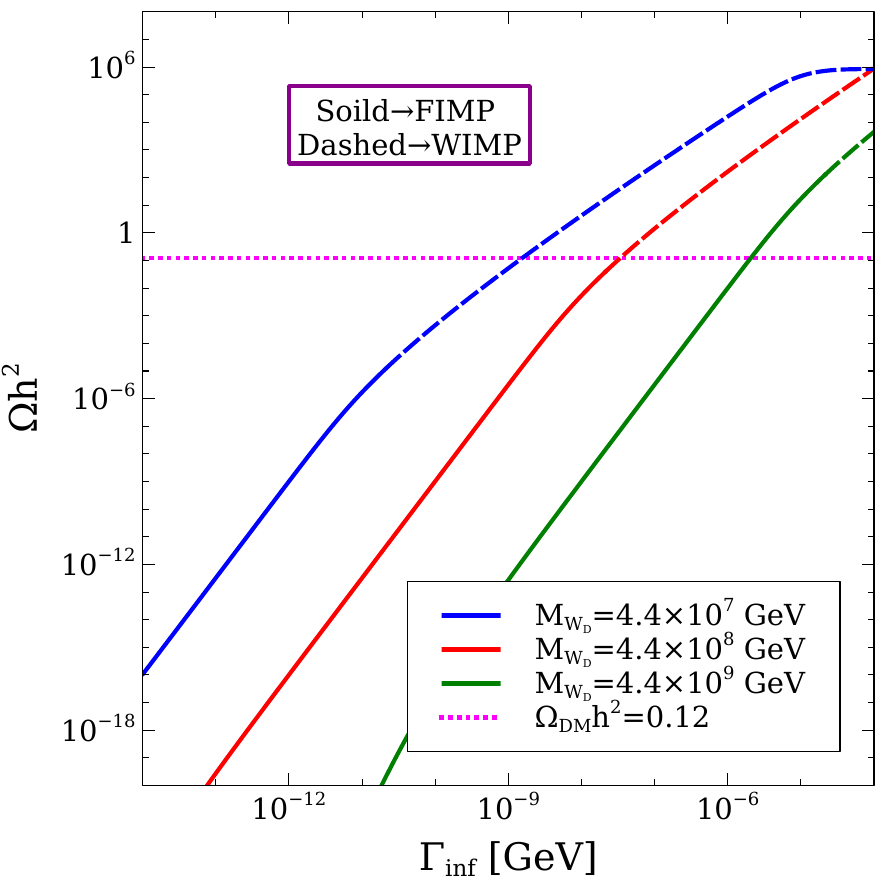}
\caption{In both plots, we show the variation of DM with the inflaton decay width. In the LP, the results are shown for three values of the gauge coupling $g_{D}$, whereas in the RP, they are shown for three values of the DM mass $M_{W_D}$.} 
\label{line-plot-2}
\end{figure}

In the LP and RP of Fig.~\ref{line-plot-2}, we show the variation of DM relic density with the inflaton decay width $\Gamma_{\rm inf}$. In the LP, results are presented for three different values of the gauge coupling $g_D$, and in the RP, for three different values of $M_{W_D}$.
From the LP, we can see that as $\Gamma_{\rm inf}$ increases, the DM relic density also increases. This is because a larger $\Gamma_{\rm inf}$ corresponds to a higher reheating temperature, which implies less dilution due to entropy production. Moreover, for $g_D=0.1$ in the LP, DM is produced via the freeze-in mechanism only and not by freeze-out. This is because the DM interaction rate, given by the product of the DM number density and $\langle \sigma v \rangle_{W_{D}W_{D}}$, is not strong enough to bring DM into thermal equilibrium following the condition in Eq. \ref{thermality-check}.
We also observe that as the gauge coupling increases, the transition from FIMP to WIMP occurs earlier. This implies that the required number density for DM to achieve thermal equilibrium is reached sooner because of the higher interaction rate.
In the RP, we show the variation for three values of DM mass $M_{W_D}$. As the DM mass increases, the FIMP-to-WIMP transition shifts toward higher values of $\Gamma_{\rm inf}$, i.e., higher reheating temperatures. This behaviour indicates that as the DM mass increases, the interaction rate decreases (as seen in Eq.~\ref{ab-psi-wd}). Therefore, to maintain DM in thermal equilibrium, a larger number density is required, which demands an earlier decay of the inflaton into the thermal bath.

\begin{figure}[h!]
\centering
\includegraphics[angle=0,height=7.5cm,width=8.5cm]{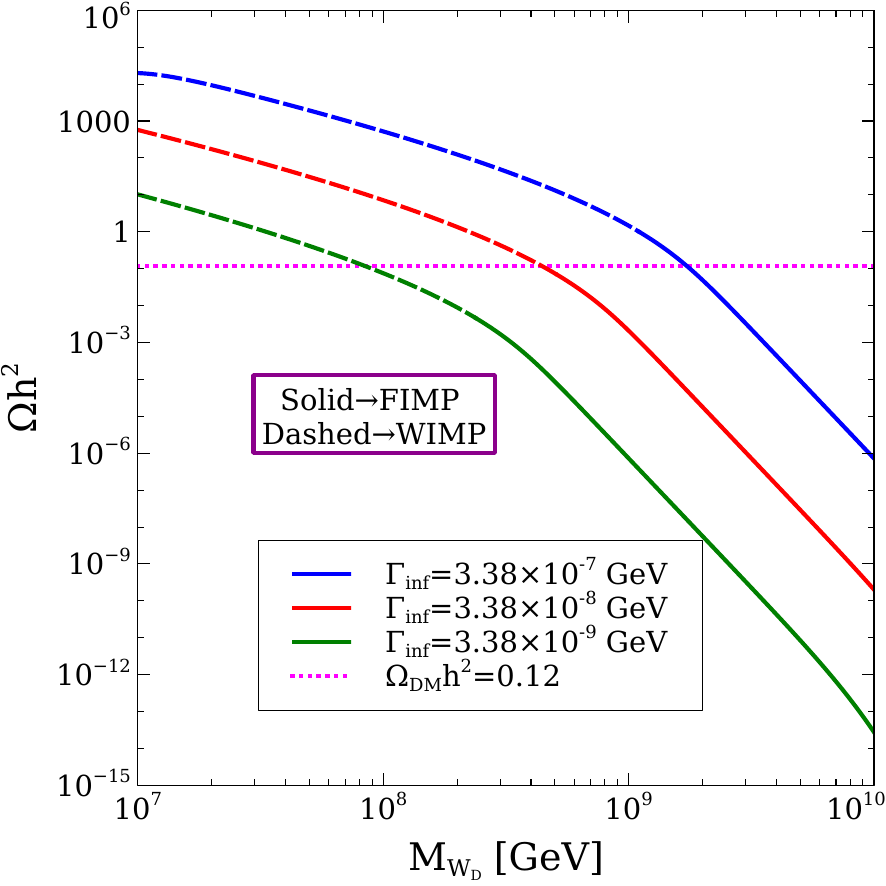}
\includegraphics[angle=0,height=7.5cm,width=8.5cm]{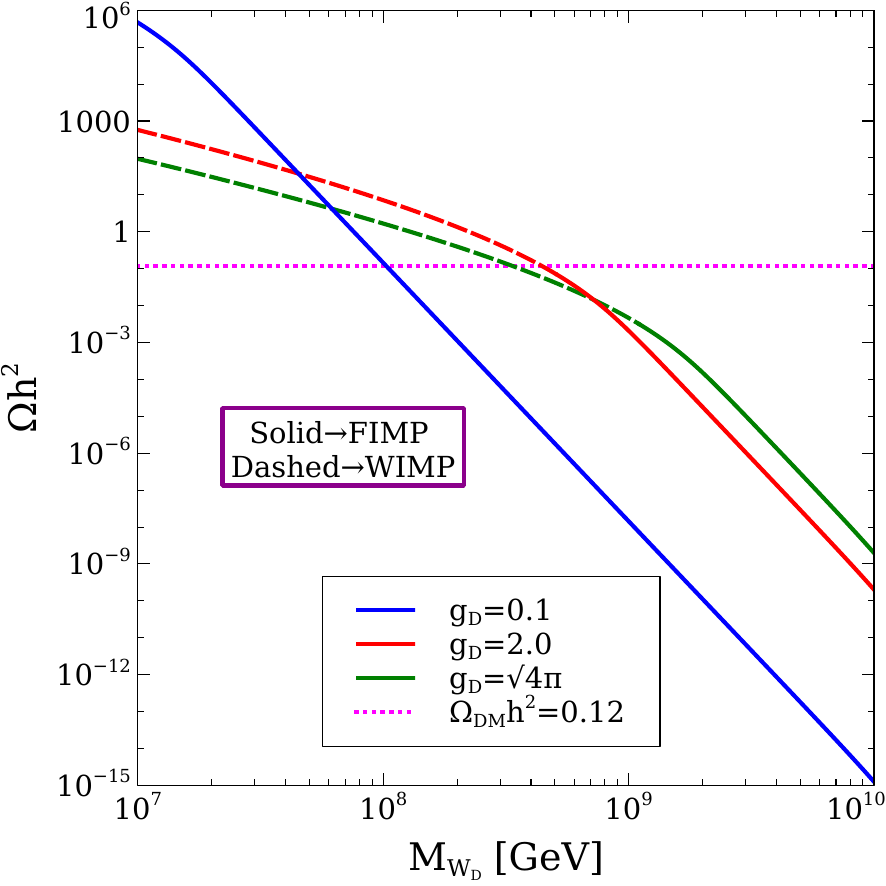}
\caption{The LP and RP show the variation of DM relic density with the DM mass for three values of the inflaton decay width $\Gamma_{\rm inf}$ and the gauge coupling $g_D$, respectively.} 
\label{line-plot-3}
\end{figure}

In the LP and RP, we show the variation of DM relic density with respect to the DM mass. In the LP, results are shown for three values of the inflaton decay width, while in the RP, they are shown for three values of the gauge coupling $g_D$.
In general, as $M_{W_D}$ increases, the cross section times velocity for $W_{D}W_{D} \longleftrightarrow h_{2} h_{2}$ decreases. This means DM freezes out earlier and therefore has more time to be diluted until the reheating temperature $T_{\rm reh}$. Consequently, we obtain a decrease in the DM relic density instead of an increase as in the usual thermal freeze-out scenario.
For the three values of $\Gamma_{\rm inf}$, we observe more DM production for larger $\Gamma_{\rm inf}$, since higher $\Gamma_{\rm inf}$ allows less time for dilution. Moreover, with increasing $\Gamma_{\rm inf}$, the WIMP-to-FIMP transition occurs at higher DM masses. This happens because the earlier production of the thermal bath from inflaton decay generates more DM, enabling it to remain in thermal equilibrium even for higher masses.
In the RP, we show the variation for three values of the gauge coupling. For $g_{D} = 0.1$, DM never thermalises, because the increased DM mass reduces the interaction rate. For $g_D = 2.0$, DM transitions from the WIMP to the FIMP regime earlier than for $g_D = \sqrt{4 \pi}$, again due to the suppression of the DM interaction with increasing mass.\\

{\bf Scatter Plots}

In the previous plots, we discussed how the DM relic density varies with different model parameters. We now turn to discuss how these parameters are correlated in order to obtain a DM relic density below the upper limit set by Planck data \cite{Planck:2018vyg}.
For generating the scatter plots, we have varied the model parameters within the following ranges,
\begin{eqnarray}
& 1 \leq \left( M_{h_2} - M_{h_1} \right)[{\rm GeV}] \leq 10^{5}, \quad 10^{7} \leq M_{W_D} [{\rm GeV}] \leq 10^{10}, \quad
10^{-5} \leq g_{D} \leq \sqrt{4 \pi}, \nonumber \\
& 10^{-4} \leq \sin\theta \leq 0.23, \quad 10^{-12} \leq \Gamma_{\rm inf}~ [{\rm GeV}] \leq 10^{-4}, \quad 10^{10} \leq H_{0}~ [{\rm GeV}] \leq 10^{11}.
\label{parameter-range}
\end{eqnarray}
The allowed points correspond to parameter choices providing the DM relic density within the range,
\begin{align}
10^{-10} \leq \Omega_{W_{D}} h^{2} \leq 0.1226.
\label{RD-limit}
\end{align}
We permit such a small DM fraction in order to obtain the KM3NeT flux, which depends on the combination $\tau_{W_D}/f_{W_D}$. For a lower DM fraction, the DM lifetime can be close to the age of the Universe. Conversely, if the DM fraction is even smaller, the required DM lifetime would be shorter than the age of the Universe, implying that the KM3NeT signal would arise primarily from the extragalactic component, as the galactic contribution would have already decayed.
This kind of tiny fraction can be easily implemented in the multicomponent DM scenario as discussed in Ref. \cite{Khan:2024biq}.

\begin{figure}[h!]
\centering
\includegraphics[angle=0,height=8.5cm,width=8.5cm]{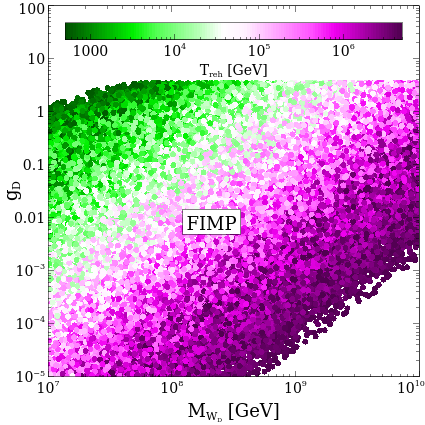}
\includegraphics[angle=0,height=8.5cm,width=8.5cm]{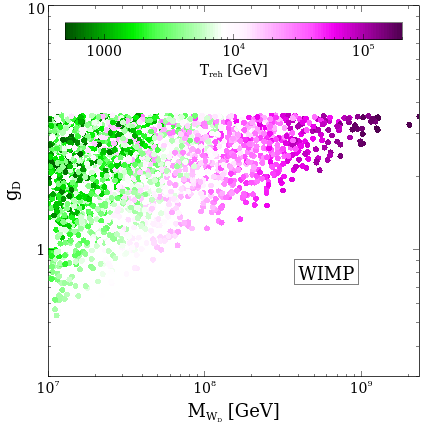}
\caption{LP and RP show the allowed regions in the $(M_{W_D},~g_{D})$ plane after varying the model parameters given in Eq. \ref{parameter-range}. The LP corresponds to FIMP-type DM, whereas the RP corresponds to WIMP-type DM.} 
\label{DM-scatter-plot-1}
\end{figure}

In the LP and RP of Fig. \ref{DM-scatter-plot-1}, we show the scatter plots in the $(M_{W_D},~g_{D})$ plane for the freeze-in and freeze-out DM production mechanisms, respectively. In both plots, the colour gradient represents the reheating temperature. The model parameters have been varied within the ranges given in Eq.\ref{parameter-range}, and the allowed data points are selected by requiring the DM relic density to lie within the range specified in Eq. \ref{RD-limit}. We have also accepted points corresponding to a very small DM fraction, since in the low-reheating scenario such a production is possible, and it also allows for a DM lifetime close to the age of the Universe. This is relevant because in the flux expression, the effective dependence is through $\tau_{W_D}/f_{W_D}$, where $f_{W_D}$ is the DM fraction.
In the LP, we observe that for a fixed gauge coupling $g_D$, increasing the DM mass requires higher values of $T_{\rm reh}$. This is because with larger $M_{W_D}$ the interaction rate decreases, leading to less DM production, and therefore a higher reheating temperature is needed to compensate for the another dilution coming from the entropy dilution. Conversely, for a fixed DM mass, increasing $g_D$ enhances DM production, which then requires stronger dilution of the DM density through entropy production, demanding lower values of $T_{\rm reh}$. The freeze-in mechanism with a low reheating scenario thus allows for a large parameter space that can explain the KM3NeT signal.
In the RP, we present the case of WIMP-type DM. Here, relatively large gauge couplings are required to obtain the correct relic density. As the DM mass increases, even larger values of $g_D$ are needed, eventually reaching the perturbative limit. Therefore, we conclude that it is challenging to obtain the correct relic density via the freeze-out mechanism for DM masses $M_{W_D} \gtrsim 10^{9}$ GeV.

\begin{figure}[h!]
\centering
\includegraphics[angle=0,height=8.5cm,width=8.5cm]{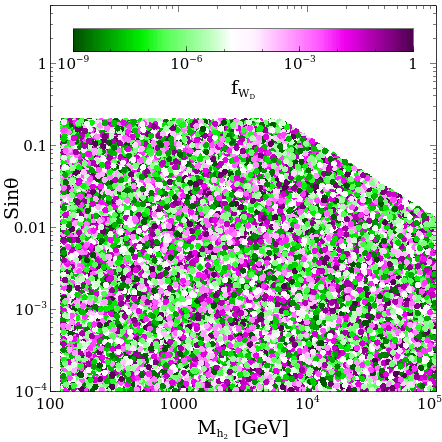}
\includegraphics[angle=0,height=8.5cm,width=8.5cm]{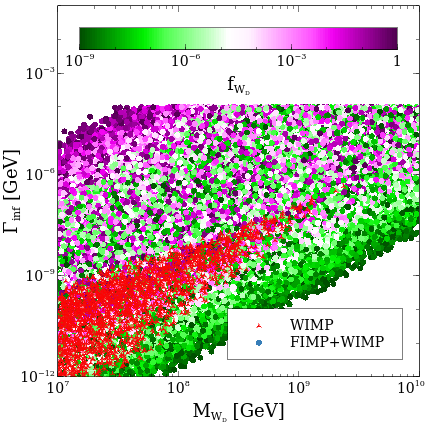}
\caption{LP shows the scatter plot in the $(M_{h_2},~\sin\theta)$ plane, whereas the RP shows the scatter plot in the $(M_{W_D},~\Gamma_{\rm inf})$ plane. In both LP and RP, we include points coming from the WIMP and FIMP type DM, with the red points region in the RP representing WIMP-type DM.} 
\label{DM-scatter-plot-2}
\end{figure}

In the LP and RP of Fig. \ref{DM-scatter-plot-2}, we show the variations in the 
$(M_{h_2},~\sin\theta)$ and $(M_{W_D},~\Gamma_{\rm inf})$ planes after imposing the DM relic density constraint discussed earlier. In both plots, the colour variation represents the fraction of DM relative to the total relic density.
In the LP, we see that the correct DM relic density can be obtained even for a TeV-scale BSM Higgs with a larger mixing angle. The top-right region is disallowed due to the violation of perturbative bounds on the quartic couplings. For most combinations of $M_{h_2}$ and $\sin\alpha$, a wide range of DM relic densities can be achieved, as considered in the present work and seen from the random colour variation. The LP includes contributions from both freeze-out and freeze-in mechanisms. We can see that the BSM Higgs mass and mixing angle are such that the BSM Higgs can be probed at the ongoing/future collider experiments.
In the RP, the outer regions are populated by magenta and green points. The magenta points correspond to 100\% DM, which arises for larger values of $\Gamma_{\rm inf}$, implying less dilution due to entropy production. The green points correspond to lower $\Gamma_{\rm inf}$ values, i.e., more dilution. Between the magenta and green points, there are mixed regions that result from different combinations of $g_D$ and $\Gamma_{\rm inf}$. The red points indicate regions where DM can only be produced via the freeze-out mechanism, in contrast to the freeze-in mechanism, where larger regions are allowed.

\section{Explaining KM3NeT Signal}
\label{KM3NeT-signal}
We can write the Lagrangian as follows, considering the mixing among the Abelian gauge symmetries:
\begin{align} \mathcal{L}_{Kinetic} &= -\frac{1}{4} \hat{B}_{\mu\nu} \hat{B}^{\mu\nu} -\frac{1}{4} \hat{W^{a}}_{\mu\nu} \hat{W^{a}}^{\mu\nu} -\frac{1}{4} \hat{W_{D}}_{\mu\nu} \hat{W_{D}}^{\mu\nu} -\frac{\alpha}{2} \hat{B}_{\mu\nu} \hat{W_{D}}^{\mu\nu} \nonumber \\ & + \frac{1}{2} \hat{M}^{2}_{\hat{Z}} \hat{Z}_{\mu} \hat{Z}^{\mu} + \frac{1}{2} \hat{M}^{2}_{W_{D}} \hat{W_{D}}_{\mu} \hat{W_{D}}^{\mu}. \label{kinetic-terms} \end{align}
In this work, we focus on the limit $\alpha \ll 1$, which guarantees that the DM lifetime is comparable to or longer than the age of the Universe.
As outlined in Refs. \cite{Heeck:2011md}, the $3\times 3$ mass matrix is written in the basis $(\hat{B}_{\mu}~~\hat{W}^{3}_{\mu}~~\hat{W}_{D\mu})$, and the kinetic terms must be diagonalised by rotating the fields. In the small kinetic mixing limit, the flavour basis $(\hat{A}~~\hat{Z}~~\hat{W}_{D \mu})$ and the mass basis $(A~~Z~~W_{D})$ are related as follows\footnote{Here $\hat{A}$ and $\hat{Z}$ are the SM photon and Z-boson, slightly shifted due to the kinetic mixing term.},
\begin{align} \begin{pmatrix} \hat{A}\\ \hat{Z}\\ \hat{W}_{D} \end{pmatrix} \simeq \begin{pmatrix} 1 & 0 & -\hat{c}_{w} \alpha \\ 0 & 1 & \frac{\hat{s}_{w} \alpha \hat{M}^2_{W_D}}{\hat{M}^2_{W_D} - \hat{M}^2_{Z}} \\ 0 & -\frac{\hat{s}_{w} \alpha \hat{M}^2_{Z}}{\hat{M}^2_{W_D} - \hat{M}^2_{Z}} & 1 \end{pmatrix} \begin{pmatrix} A\\ Z\\ W_{D} \end{pmatrix} \end{align}
The masses in the flavour and mass bases are related as,
\begin{align} 
M^2_{Z} \simeq \hat{M}^2_{\hat{Z}} + \frac{\hat{M}^4_{\hat{Z}} \hat{s}^2_{w} \alpha^2}{\hat{M}^2_{\hat{Z}} - \hat{M}^2_{W_D}},~~ M^2_{W_D} \simeq \hat{M}^2_{\hat{W}_D} - \frac{\hat{M}^4_{\hat{Z}} \hat{s}^2_{w} \alpha^2}{\hat{M}^2_{\hat{Z}} - \hat{M}^2_{W_D}}, 
\label{mass-releation}
\end{align}
Since $\alpha$ is a tiny parameter, the masses can be treated as effectively identical in both bases.

\begin{figure}[h!]
\centering
\includegraphics[angle=0,height=8.5cm,width=8.5cm]{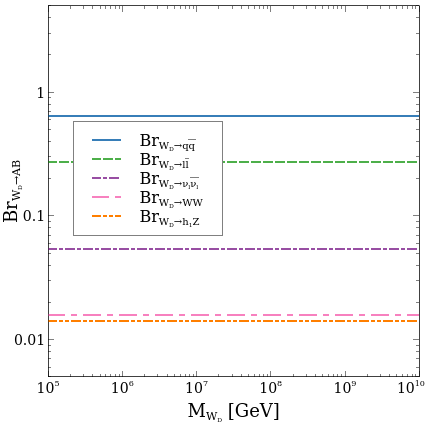}
\includegraphics[angle=0,height=8.5cm,width=8.5cm]{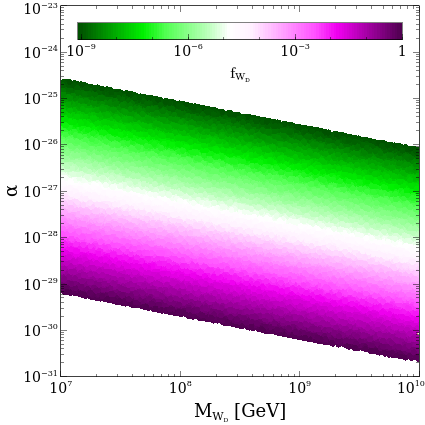}
\caption{LP shows the decay branching ratios of the vector DM due to the kinetic mixing between the dark abelian gauge symmetry and the hypercharge gauge group. The DM predominantly decays into $q\bar{q}$ modes. In the RP, we show the variation in the $(M_{W_D},~\alpha)$ plane, where the colour variation represents the DM fraction. In producing the RP, we have fixed the DM lifetime at $\tau_{W_D} = 10^{29}$ sec.} 
\label{km3net-plot-1}
\end{figure}

As mentioned earlier, in this work we explain the KM3NeT signal through DM decay without relying on astrophysical sources. 
The KM3NeT collaboration \cite{KM3NeT:2025npi} has also confirmed that the absence of any known astrophysical sources along the direction of the signal.
One advantage of explaining the KM3NeT signal via DM decay is that, by appropriately choosing the DM mass, lifetime, and fraction of the  DM component, the signal can be easily accommodated. In this study, we focus on vector--type DM decay, which predominantly produces neutrinos along with other SM particles.
The DM decay has two dominant contributions, one from the Galactic component and the other from the extragalactic component. As described in Ref. \cite{KM3NeT:2025npi}, the signal originates from a direction away from the GC in the equatorial coordinate system\footnote{In the galactic coordinate system $(l,b) = (210.06^{\circ}, -11.13^{\circ})$}, RA: $94.3^{\circ}$, Dec: $-7.8^{\circ}$, with a $1\sigma$ angular uncertainty of $\pm 1.5\sigma$. 
Since this direction is away from the GC, our results are largely insensitive to the choice of DM distribution near the GC.
We adopt the widely used NFW profile for DM distribution in the galaxy \cite{Navarro:1995iw},
\begin{eqnarray}
\rho_{DM}(r) = \frac{\rho^0_{DM}}{\frac{r}{r_{c}} \biggl( 1 + \frac{r}{r_{c}} \biggr)^{2}},
\label{NFW-profile}
\end{eqnarray}
where $r_{c} \simeq 20$ kpc, $\rho^{0}_{DM} \simeq 0.3$ GeV/cm$^{3}$, and $r$ is the distance from the GC. The Galactic contribution to the neutrino flux from DM decay can be parameterised as follows,
\begin{eqnarray}
\frac{d\phi^G_{\nu}}{d E_{\nu} d\Omega}\bigg|_{E_{\nu} = E} = \frac{D_{G}}{4 \pi W_{D}}
\sum_{i} \Gamma_{i} \frac{d N^i_{\nu}}{dE_{\nu}}\bigg|_{E_{\nu} = E},
\end{eqnarray}
where $\Gamma_{i}$ is the DM decay width for $i = q\bar{q},l \bar{l},\nu_{l} \bar{\nu}_l,~W^{+} W^{-},~h_{1} Z$, and their branching ratios are given in the LP of Fig. \ref{km3net-plot-1}. $\frac{d N^{i}_{\nu}}{d E_{\nu}}$ is the neutrino spectrum as a function of energy, generated using the HDMSpectra package \cite{Bauer:2020jay}. Finally, we sum over all the channels.
The factor $D_G$ is the D-factor, which depends on the DM distribution in our galaxy and the angular direction of the signal, and is defined as follows,
\begin{eqnarray}
D_{G} = \frac{f_{W_D}}{\Delta\Omega} \int_{\Delta \Omega} d\Omega \int_{0}^{r_{max}} dr^{\prime} \rho_{DM}(r^{\prime})
\end{eqnarray}
where $f_{W_D} = \frac{\Omega_{W_D} h^{2}}{0.12}$, $\Delta\Omega$ is the $1\sigma$ angular variation along the direction of the event. The DM density $\rho_{W_D}(r^{\prime})$ is defined in Eq. \ref{NFW-profile}, and $r^{\prime} = \sqrt{r^2_{\odot} + r^{2} - 2 r_{\odot} r \cos\psi}$, where $r_{\odot} = 8.5$ kpc is the distance between the GC and the Solar System, $r$ is the distance between the DM decay point and the Earth, and $\psi$ is the angular separation.
The other contribution to the neutrino flux comes from the extragalactic component of DM decay. In the extragalactic contribution, one must account for the redshifting of energy due to the expansion of the Universe, {\it i.e.}, when DM decays at redshift $z$. The neutrino flux from the extragalactic contribution can be written as follows \cite{Cirelli:2010xx},
\begin{align}
\frac{d \phi^{EG}_{\nu}}{d E_{\nu} d \Omega} = \int_{z_{0}}^{\infty} dz \frac{1}{\mathcal{H}(z)} \biggl( \frac{1+z_{0}}{1+z} \biggr)^{3} \frac{\bar \rho_{W_D}}{M_{W_D}} \sum_{i} \Gamma_{i} \frac{d N^i_{\nu}}{d E^{\prime}_{\nu}}\bigg|_{E^{\prime}_{\nu} = (1+z) E_{\nu}} e^{-S_{\nu}\left(E_{\nu},z\right)} 
\end{align}  
where $\bigl( \mathcal{H}(z) (1+z) \bigr)^{-1}$ represents the proper distance, taking into account the redshift, $\bigl( \frac{1+z_{0}}{1+z} \bigr)$ accounts for the dimming of surface brightness due to the dilution of sources from the expansion of the Universe. Here, $z_0 = 0$ is the redshift today, and $z$ is the redshift at the time of DM decay\footnote{For example, $z=3200$ corresponds to the epoch of matter--radiation equality, below this value, matter domination occurs \cite{Planck:2018vyg}.}. The average cosmological DM density is $\bar \rho_{W_D}(z) = f_{W_D} \rho^0_{DM} (1+z)^{3}$.

The Hubble parameter $\mathcal{H}(z)$, valid for all eras of the Universe at redshift $z$, is expressed as
\begin{eqnarray}
\mathcal{H}(z) = H_{0} \sqrt{\Omega_{\Lambda} + \Omega_{m} (1+z)^{3} + \Omega_{r} (1+z)^{4}},
\end{eqnarray}
where $H_{0} = 67.4~{\rm km  s^{-1} Mpc^{-1}}$ is the Hubble parameter today \cite{Planck:2018vyg}.
The neutrino flux observed today at redshift $z_0 = 0$ with energy $E$, originating from emission at redshift $z$ with energy $E^{\prime}_{\nu} = (1+z)E_{\nu}$, can be expressed as
\begin{eqnarray}
\frac{d \phi^{\rm EG}_{\nu}}{d E_{\nu} d\Omega} = D_{\rm EG} \int_{0}^{\infty} dz \frac{1}{\sqrt{\Omega_{\Lambda} + \Omega_{m} (1+z)^{3} + \Omega_{r} (1+z)^{4}}} \frac{dN^{i}_{\nu}}{d E^{\prime}_{\nu}}\bigg|_{E^{\prime}_{\nu} = (1+z) E_{\nu}} e^{-S_{\nu}(E_{\nu},z)}.
\end{eqnarray}

The D-factor for the extragalactic contribution, which depends on the DM mass, lifetime, and fraction, is given by
\begin{align}
D_{\rm EG} = \frac{f_{W_D} \Omega_{\rm DM} \rho_{c}}{4 \pi M_{W_D} \tau_{W_D} H_{0}}
= 3.64 \times 10^{-17} \biggl( \frac{440~{\rm PeV}}{M_{W_D}} \biggr) \biggl( \frac{10^{29}{\rm s}}{\tau_{W_D}/f_{W_D}} \biggr){\rm cm^{2} s^{-1} sr^{-1}},
\end{align}
where the critical density $\rho_c = \frac{3 H^2_{0}}{8 \pi G} = 5.5 \times 10^{-6}~{\rm GeV cm^{-3}}$, $\Omega_{\rm DM} = 0.27$ is the DM fraction, $\Omega_m = 0.3147$ is the total matter fraction, $\Omega_\Lambda = 0.6847$ is the cosmological constant fraction, and $\Omega_r = \Omega_m / (1+z_{\rm eq})$ is the radiation fraction, taken from Planck 2018 data \cite{Planck:2018vyg}.
Finally, we include the neutrino opacity factor given by \cite{Esmaili:2012us},
\begin{eqnarray}
S_{\nu}(E_{\nu},z) =
\begin{cases}
7.4\times 10^{-17} (1+z)^{7/2} \biggl( \frac{E_{\nu}}{\rm TeV} \biggr), & {\rm for} 1 \ll z < z_{\rm eq}, \\
1.7\times 10^{-14} (1+z)^{3} \biggl( \frac{E_{\nu}}{\rm TeV} \biggr), & {\rm for} z \gg z_{\rm eq},
\end{cases}
\end{eqnarray}
where $z_{\rm eq} = 3208$ is the redshift at matter--radiation equality, as defined earlier \cite{Planck:2018vyg}.

\begin{figure}[h!]
\centering
\includegraphics[angle=0,height=8.5cm,width=13.5cm]{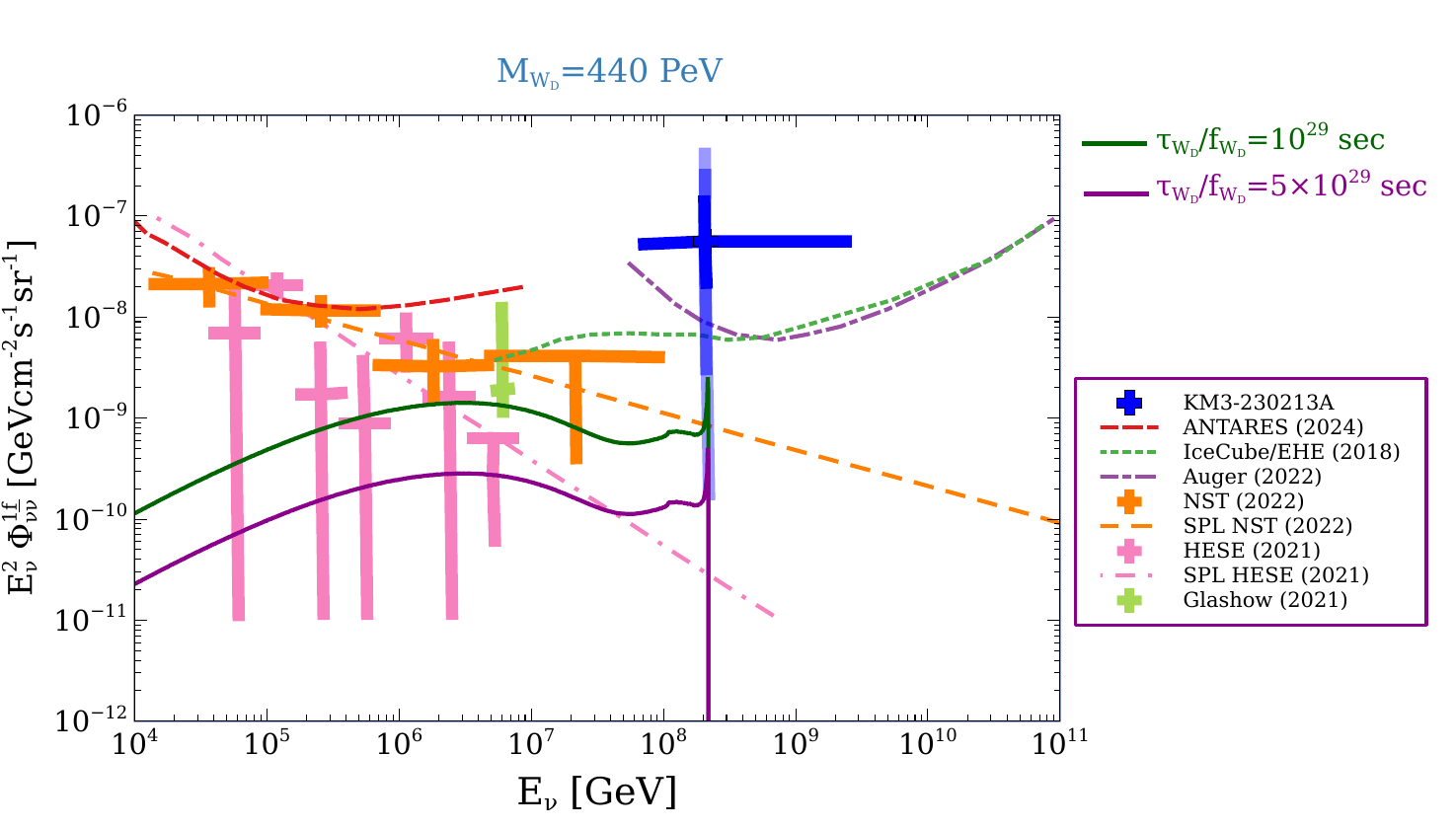}
\caption{Variation of the neutrino flux originating from DM decay with energy for two different values of the DM decay lifetime. Here, the neutrino flux includes both galactic and extragalactic contributions. The flux also explains part of the IceCube data, with a peak at the KM3NeT data point. The magenta points correspond to IceCube-HESE data \cite{IceCube:2020wum}, the orange points are from IceCube-NST data \cite{Abbasi:2021qfz}, the red dashed line represents the ANTARES limit \cite{ANTARES:2024ihw}, the blue point is the KM3NeT data \cite{KM3NeT:2025npi}, the green dashed line is the limit from IceCube \cite{IceCube:2018fhm}, and the magenta dashed-dotted line is the limit from the PAO data \cite{PierreAuger:2023pjg}.} 
\label{km3net-plot-2}
\end{figure}

In the LP of Fig.\ref{km3net-plot-1}, we show the branching ratios of $W_D$ to different channels. The analytical expressions of the $W_D$ decay modes are displayed in section \ref{WD-dec-expression}. We see that the dominant decay mode is the $q\bar{q}$ channel, followed by $l \bar{l}$ and $\nu_l \bar{\nu}_l$, with relatively smaller contributions from the $WW$ and $h_{1}Z$ decay modes. The branching ratios are well-suited for explaining the KM3NeT signal. In general, the $q\bar{q}$ and $l\bar{l}$ decay modes contribute similarly to the photon and neutrino fluxes, but we also have a direct decay to neutrinos, which can peak at the KM3NeT signal point and produce the desired flux.
On the other hand, in the RP, we show scatter plots in the $(M_{W_D},~\alpha)$ plane, with the colour variation representing the fraction of DM compared to the total DM density. In generating this plot, we have fixed the DM lifetime at $\tau_{W_D} = 10^{29}$ sec, therefore, the KM3NeT signal can be explained for a DM mass around $M_{W_D} \sim 4.4 \times 10^{8}$ GeV.

In Fig.~\ref{km3net-plot-2}, we show the neutrino flux obtained in our work from DM decay. We assume that the neutrino flux, consisting of all flavours from DM decays, is in the $1:1:1$ ratio. In the plot, the green and magenta solid lines are generated from DM decay for lifetimes $\tau_{W_D}/f_{W_D} = 10^{29}$ sec and $5\times10^{29}$ sec, respectively.
We see that the flux obtained from DM decay is below the limit set by non-observation at IceCube \cite{IceCube:2018fhm}. This is an advantage of the decaying DM scenario, which can explain the KM3NeT signal without violating the IceCube limit. We have considered a DM mass $M_{W_D} = 4.4 \times 10^{8}$ GeV, which helps to peak the neutrino flux at the central value. The DM decay lifetime can be chosen in the range $\tau_{W_D}/f_{W_D} = 5 \times 10^{28} - 10^{30}$ sec to explain the KM3-230213A event.
In the plot, the magenta points correspond to IceCube-HESE data \cite{IceCube:2020wum} and the orange points correspond to IceCube-NST data \cite{Abbasi:2021qfz}. The red dashed line represents the ANTARES limit from non-observation of neutrinos \cite{ANTARES:2024ihw}, the green dashed line is the IceCube limit for non-observation of neutrinos at these energies \cite{IceCube:2018fhm}, and the magenta dashed-dotted line corresponds to the PAO limit \cite{PierreAuger:2023pjg}. The Glashow resonance point is shown by the light green point. We see that our neutrino flux can explain part of the observed IceCube neutrino data and the Glashow resonance point for suitable values of the DM decay lifetime.

\begin{figure}[h!]
\centering
\includegraphics[angle=0,height=8.5cm,width=11.0cm]{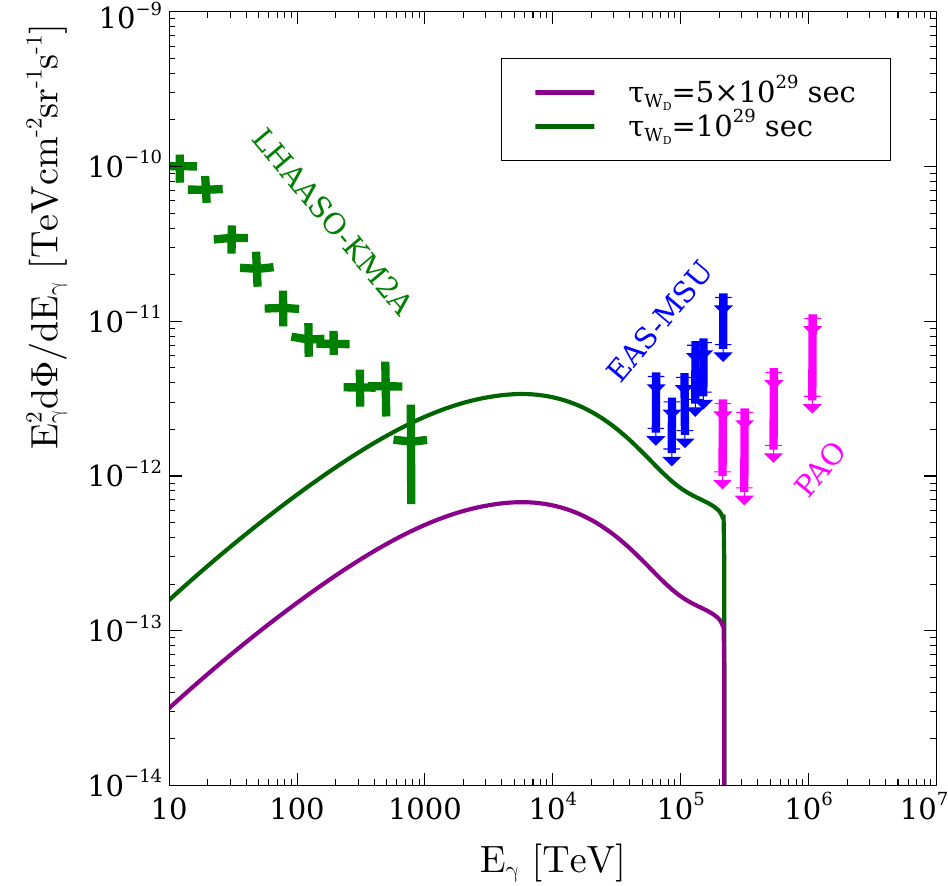}
\caption{Variation of the photon flux originating from DM decay with energy. The flux includes only the galactic contribution, while the extragalactic contribution scatters with the background neutrinos and is shifted towards lower energy \cite{Jho:2025gaf}. For lower values of the DM decay lifetime, the flux approaches the LHAASO-KM2A data \cite{LHAASO:2023gne}.} 
\label{km3net-plot-3}
\end{figure}

In Fig. \ref{km3net-plot-3}, we show the photon flux obtained from DM decay. In the present work, for the photon flux, we have considered only the galactic contribution and not the extragalactic contribution, because for $\mathcal{O}(\text{PeV})$-scale photons, they scatter with background photons, resulting in a mean free path of less than 1 Mpc \cite{Murase:2010va, Murase:2012xs}. Moreover, PeV-scale photons undergo pair annihilation with background photons and are subsequently reproduced through Compton scattering at lower energies, which shifts the photon spectrum to lower energies \cite{Jho:2025gaf}.
In the figure, the green points correspond to LHAASO-KM2A data \cite{LHAASO:2023gne}, whereas the blue and magenta points indicate limits from the EAS-MSU \cite{Fomin:2017ypo} and PAO \cite{Castellina:2019huz} experiments. The photon flux does not exhibit a peak like the neutrino flux because there is no direct production of photons from DM, but it has a broader spectrum ranging from 10 TeV to $2\times10^{5}$ TeV. The green and magenta solid lines represent the photon flux obtained from DM decay in our work, with DM lifetimes $\tau_{W_D} = 10^{29}$ sec and $5\times 10^{29}$ sec, respectively. In generating the flux, we have considered the same inner galactic plane as in the LHAASO-KM2A experiment, namely $15^{\circ} < l < 135^{\circ}$ and $-5^{\circ} < b < 5^{\circ}$.
We can see that part of the flux reaches the LHAASO-KM2A data points, but it is unable to explain the entire LHAASO-KM2A dataset.

\section{Gravitational Waves}
\label{sec:GW}
In the present work, we have considered $U(1)_{D}$ gauge symmetry, which spontaneously breaks when the BSM scalar acquires a vev. This breaking of the abelian gauge symmetry $U(1)_D$ can form cosmic strings in the early Universe, with string tension proportional to the square of the vev, {\it i.e.} $\mu \sim v^2_{D}$. During the evolution of the Universe, cosmic strings (CS) not only stretch but can intercommute, forming loops that radiate energy in the form of GWs. 
Here we mainly focus on cosmic strings that significantly produce GWs through loop formation, in contrast to Refs.~\cite{Vincent:1997cx, Bevis:2006mj, Figueroa:2012kw} where local CS are argued to decay dominantly to massive vector bosons and Higgs quanta. Additionally, Refs.~\cite{Srednicki:1986xg, Vilenkin:1986ku, Damour:1996pv} show that global strings dominantly decay to light Goldstone quanta.

We focus on CS that decay dominantly to GWs. Due to GW emission, CS never dominate the Universe energy budget, even though their energy density redshifts as $\rho_{\rm string} \propto a^{-2}$. This balance between energy emission and $a^{-2}$ redshift of the energy density leads the cosmic string energy density to reach a fraction of the total energy, known as the scaling regime \cite{Albrecht:1984xv, Bennett:1987vf, Allen:1990tv, Zeldovich:1978wj}. The GW production from CS in the scaling regime can be estimated analytically using the velocity-dependent one-scale (VOS) model \cite{Martins:1995tg, Martins:1996jp, Martins:2000cs, Avelino:2012qy, Sousa:2013aaa}, where the string length and velocity are taken into account and match well with the simulation results. As found in Refs.~\cite{Vanchurin:2005pa, Martins:2005es, Olum:2006ix, Ringeval:2005kr, Blanco-Pillado:2011egf, Blanco-Pillado:2017oxo} from simulations, during loop formation from long cosmic strings, only $10\%$ form large loops that primarily produce GWs, while the remaining $90\%$ are small loops that dilute simply due to redshifting. In the VOS model, with initial string size $l_i = \alpha t_i$, the loop formation rate considering a fraction $F_{\alpha}$ of energy going into large loops is,
\begin{align}
\frac{dn_{\alpha}}{dt} = F_{\alpha} \frac{C_{\rm eff}^{\rm phase}}{\alpha} t^{-4},
\end{align} 
where $F_{\alpha} \simeq 0.1$, $\alpha = 0.1$, $C^{\rm mat}_{\rm eff} \simeq 0.39$ during matter domination era, and $C^{\rm rad}_{\rm eff} \simeq 5.4$ during radiation domination era \cite{Blanco-Pillado:2011egf, Blanco-Pillado:2013qja, Blanco-Pillado:2013qja}. 
Once large loops are formed, they emit GWs at a constant rate as,
\begin{align}
\frac{dE}{dt} = - \Gamma G \mu^{2},
\end{align}
where $\Gamma \simeq 50$. Solving this equation with initial energy $E_i = \mu l_i$, the loop length at time $t$ is,
\begin{align}
l(t) = l_i - \Gamma G \mu (t - t_i).
\end{align}
The energy loss can be decomposed into normal oscillation modes $k$ with frequency $\tilde{f}_k = \frac{2k}{l}$, where $k = 1,2,3...~$. Refs. \cite{Blanco-Pillado:2013qja, Blanco-Pillado:2013qja} show that the GW power in each mode decreases as $k^{-4/3}$, so for each mode:
\begin{align}
\Gamma^k = \frac{\Gamma k^{-4/3}}{\sum^{\infty}_{m=1} m^{-4/3}}.
\end{align}
The GW emitted at time $\tilde{t}$ with frequency $\tilde{f}$ redshifts today to frequency $f$ as,
\begin{align}
\tilde{f} = \frac{a(t_0)}{a(\tilde{t})} f. 
\end{align}
Combining the loop formation rate and GW emission, the relic GW density can be expressed as,
\begin{align}
\Omega_{\rm GW} = \frac{f}{\rho_c} \frac{d \rho_{\rm GW}}{df},
\end{align}
where $f$ is the GW frequency today, and $\rho_{\rm GW}$ is the energy density of GWs. Summing over all modes gives the total GW spectrum \cite{Cui:2018rwi},
\begin{align}
\Omega_{\rm GW}(f) = \sum_{k} \Omega_{\rm GW}^{(k)}(f),
\end{align}
where
\begin{align}
\Omega_{\rm GW}^{(k)}(f) = \frac{1}{\rho_c} \frac{2 k}{f} \frac{F_{\alpha} \Gamma^{k} G \mu^{2}}{\alpha (\alpha + \Gamma G \mu)} 
\int_{t_F}^{t_0} \frac{C_{\rm eff}^{\rm phase} (t_i^{(k)})}{(t_i^{(k)})^{4}} 
\left[\frac{a(\tilde{t})}{a(t_0)}\right]^{5} 
\left[\frac{a(t_i^{(k)})}{a(\tilde{t})}\right] 
\theta(t_i^{(k)} - t_F) dt,
\end{align}
with
\begin{align}
t_i^{(k)}(\tilde{t}, f) = \frac{1}{\alpha + \Gamma G \mu} \left[ \frac{2 k}{f} \frac{a(\tilde{t})}{a(t_0)} + \Gamma G \mu \tilde{t} \right],
\end{align}
where $t_F$ is the time when the scaling regime is achieved, and $t_0$ is the present time. Other parameters have been defined in previous sections.
As considered in section \ref{LRH-WIMP-FIMP}, the inflaton field evolves like a matter field with potential $V(\phi) = \lambda_{\phi} M_{\rm pl}^2 \phi^2$ ($\phi \ll M_{\rm pl}$) and continuously decays into the SM bath. Therefore, during reheating and before the reheating temperature, the Universe temperature varies with scale factor as $T \propto a^{-3/8}$ \cite{Garcia:2020wiy}. After reheating ends, we calculate the scale factor using the built-in relativistic degrees of freedom in \texttt{micrOMEGAs} (v6.2.3). 
On top of the cosmic-string contribution to the GW spectrum, there is also a contribution to GWs induced by primordial scalar perturbations at second order. In the present work, we have considered the gradual decay of the inflaton into the radiation bath during the early matter-dominated epoch, which further suppresses this kind of GW contribution from the early Universe, as found in Ref.~\cite{Inomata:2019zqy}. Moreover, in Ref.~\cite{Inomata:2019ivs}, the authors have pointed out a scenario based on the sudden-decay approximation, in which the inflaton does not decay into radiation until the transition time from the early matter-dominated era to the radiation era. During this transition, the inflaton suddenly decays into the radiation bath. Such a sudden transition can enhance the GW contribution because the scalar modes oscillate rapidly well inside the Hubble horizon, and the gravitational potential does not have enough time to decay. In the same reference, it has been shown that GWs generated from such a sudden transition can be detectable in proposed future detectors when the reheat temperature considered in the range $20\,{\rm GeV}\leq T_{reh} \leq 2\times 10^{7}\,{\rm GeV}$.
To realise such a sudden transition, one needs to consider an intermediate ‘triggeron’ field that blocks the decay of the inflaton until the matter-to-radiation transition time. Due to the dynamical evolution of this intermediate field, the inflaton decay into the radiation bath becomes kinematically allowed during the transition and occurs suddenly. In the present work, we have not considered such a sudden decay, instead, we have assumed a gradual decay of the inflaton. Hence, the GW contribution sourced by adiabatic perturbations generated during inflation can be ignored. 

\begin{table}[h!]
\centering
\begin{tabular}{|c|c|c|c|c|c|c|c|}
\hline
DM Nature & $f_{W_{D}} = \frac{\Omega_{W_{D}}h^{2}}{\Omega^{tot}_{DM}h^{2}}$ & $M_{h_2}$ [TeV] & $G\mu$  & $g_{D}$ & $\sin\alpha$ & $\Gamma_{\rm inf}$ [GeV] & $T_{\rm reh}$ [GeV] \\
\hline
FIMP & $8.86\times10^{-13}$ & $10^{4}$ & $1.3\times10^{-11}$ & $10^{-5}$ & $0.1$ & $10^{-11}$ & $1.71 \times 10^{3}$ \\
\hline
WIMP & $4.19\times 10^{-7}$ & $10^{4}$ & $1.3\times10^{-17}$ & $10^{-2}$ & $0.01$ & $10^{-10}$ & $5.38\times10^{3}$ \\
\hline
FIMP & $3.75\times10^{-12}$ & 3 & $1.3\times10^{-19}$ & 0.1 & 0.1 & $3.38\times10^{-10}$ & $9.88\times 10^{3}$ \\
\hline
WIMP & $4.91\times10^{-12}$ & $5\times10^{3}$ & $1.3\times10^{-17}$ & $10^{-2}$ & 0.05 & $10^{-12}$ & $5.38\times 10^{2}$ \\
\hline
FIMP & 1 & 0.3 & $3.25\times10^{-22}$ & 2.0 & 0.1 & $3.38\times10^{-8}$ & $9.88\times10^{4}$ \\
\hline
\end{tabular}
\caption{Benchmark points for obtaining the DM relic density and the corresponding parameter values for the GW study. The DM mass has been kept fixed at $M_{W_D} = 440$ PeV. }
\label{tab-GW}
\end{table}

\begin{figure}[h!]
\centering
\includegraphics[angle=0,height=8.5cm,width=12.5cm]{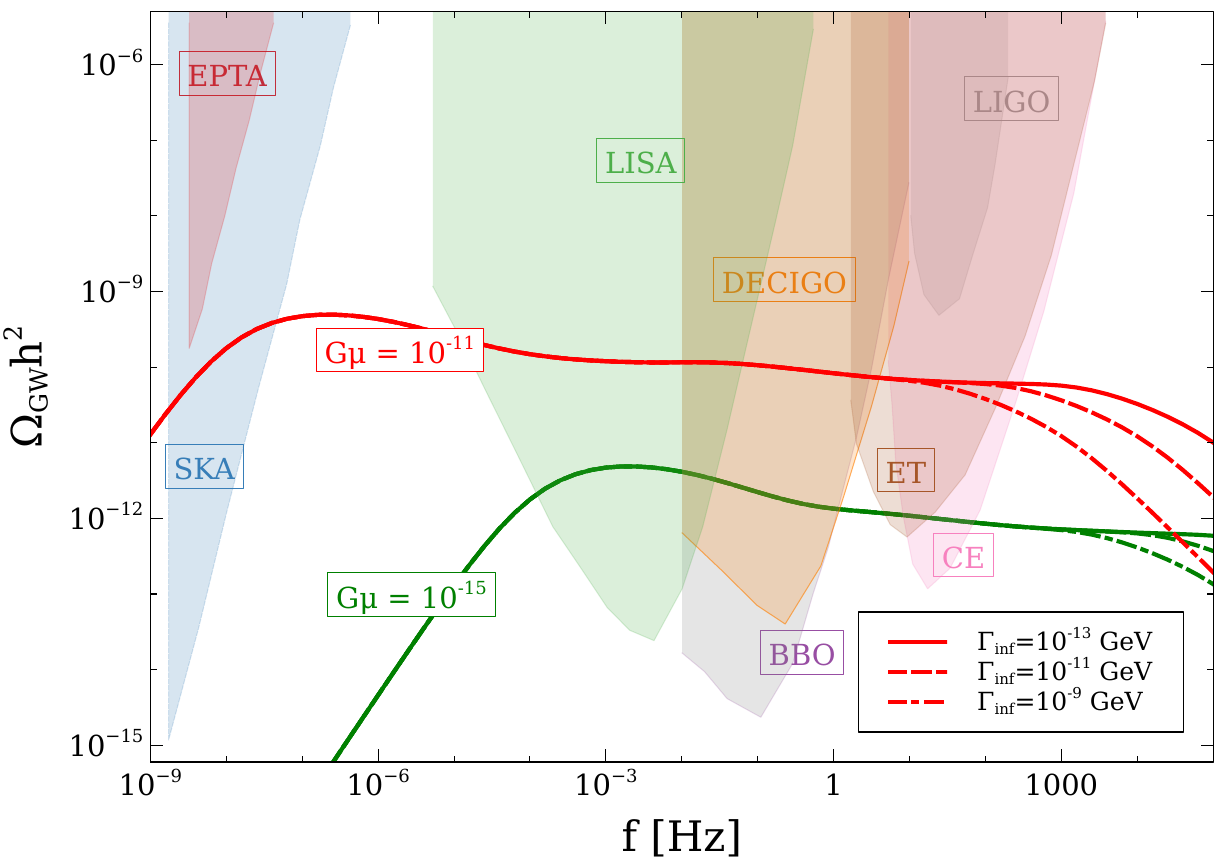}
\caption{GW density variation as a function of frequency for two different values of $G\mu$. Additionally, the plot illustrates the effect of the reheating temperature when the inflaton behaves as a matter-like field.} 
\label{GW-plot}
\end{figure}

In Fig. \ref{GW-plot}, we show the GW relic density as a function of frequency, originating from cosmic strings. In Table \ref{tab-GW}, we present a few benchmark points that generate a suitable fraction of the DM relic density capable of explaining the KM3NeT signal, while also providing values of $G\mu$ detectable by future proposed GW detectors. In the plot, we display results for two values, $G\mu = 10^{-11}$ and $10^{-15}$. There exists a bound $G\mu < 2 \times 10^{-11}$ from pulsar timing data due to the non-observation of stochastic GWs. We can see that both values fall within the sensitivity of most proposed GW detectors. For frequencies $f > 100$ Hz, there is a drop in the GW density due to the effects of the low reheating scenario. This occurs because the temperature and scale factor evolve as $a \propto T^{-3/8}$ during the reheating era, rather than $a \propto T^{-1}$ as in the radiation-dominated era.

\section{conclusion}
\label{conclusion}
In the present work, we have explained the recently observed KM3-230213A event by considering the decaying SHDM. For explaining the signal, we need a DM which dominantly decays to neutrinos with the mass in the range $M_{DM} = (1.5-52)\times 10^{8}$ GeV and lifetime $\tau_{DM} = (5.4-14.2) \times 10^{29}$ sec \cite{Kohri:2025bsn}. In this context, we have extended the SM gauge sector and particle content by dark $U(1)_D$ abelian gauge group and a SM singlet scalar charged under $U(1)_D$. The gauge boson coming from $U(1)_D$ can be a suitable DM candidate whose mass is obtained when the singlet scalar takes the vev.

In producing the SHDM, the normal freeze-out mechanism fails due to overproduction, and the freeze-in mechanism demands a very tiny gauge coupling value. To overcome these, we need to reduce the DM abundance and one of the mechanisms is through the entropy injection when the inflaton decays to the SM bath in a low reheating scenario. Due to the continuous entropy injection, we have achieved the correct value of DM relic density both for freeze-out and freeze-in mechanisms at a reasonable gauge coupling value. Depending on the low reheating temperature {\it i.e.} the inflaton decay width, we can also have DM abundance in a tiny fraction, which will demand the DM lifetime to be shorter than the age of the Universe to explain the KM3NeT flux. In that case, we only need the redshifted extragalactic contribution to explain the signal, as SHDM in our galaxy has already decayed.
This kind of scenario can be appealing because the direction of the signal is away from the GC, but we have considered both galactic and extragalactic contributions.

Moreover, to adopt the decaying DM scenario, we have utilised the kinetic mixing term with the hypercharge gauge group which allows DM to decay into SM particles. We have found that the dominant decay modes include quarks, leptons, and neutrinos, which help us to explain the KM3NeT signal with a suitable value of the DM mass, lifetime and its fraction compared to the total DM abundance. It is to be noted that the direct decay of DM to $WW$ is suppressed by the mutual cancellation of the mixing parameters, which suppresses the $M^5_{W_D}$ dependence in the decay width. We have shown the impact of different model parameters on DM production and presented a few scatter plots which show the correlation among the model parameters after demanding the upper limit of the DM relic density bound. We have found that the BSM Higgs mass and the mixing angle can be in the collider accessible range, and they do not affect the DM abundance significantly.
Depending on the choice of the gauge coupling, DM mass, and low reheating temperature, we can suitably obtain the desired value of decay lifetime by varying the kinetic mixing parameter to generate the expected neutrino flux for KM3NeT signal by obeying the upper limit set by IceCube.

Finally, our $U(1)_D$ breaks spontaneously, which can produce cosmic strings in the early Universe. Since our DM mass is super-heavy, we have a large singlet scalar vev which provides larger string tension, and as a result, we expect GW signal to be detectable at future GW observatories, including SKA, LISA, DECIGO, BBO, ET, and CE.
Additionally, the Universe up to the reheating temperature is dominated by the inflaton, the scale factor and temperature vary as $a \propto T^{-3/8}$ instead of $a \propto T^{-1}$ as in the radiation era. This different relation between $a$ and $T$ suppresses the GW spectrum at higher frequencies depending on the reheating temperature. The detection of such peculiar behaviour of GW in the future GW detectors will confirm the presence of non-standard Universe evolution in the early Universe. Therefore, our work can explain the KM3NeT signal with a proper production mechanism for SHDM, which has an impact on the GW in a low reheating scenario, and the detection of such a GW spectrum hints at the presence of non-standard evolution of the Universe in the early times. 

\section{Acknowledgements}
The authors thank Julian Heeck for email communication on Ref. \cite{Heeck:2011md}, and Alexander Pukhov for queries related to micrOMEGAs.
The research was supported  in part by Brain Pool program funded by the Ministry of Science and ICT through the National Research Foundation of Korea (RS-2024-00407977 (SK) ) and Basic Science
Research Program through the National Research Foundation of Korea (NRF) funded by the Ministry of Education, Science and Technology (NRF-2022R1A2C2003567 and RS-2024-00341419 (JK)). 
This research was supported  in part by the Excellence Cluster ORIGINS which is funded by the Deutsche Forschungsgemeinschaft (DFG, German Research Foundation) under Germany's Excellence Strategy–EXC-2094-390783311.
For the numerical analysis, we have used the Scientific Compute Cluster at GWDG, the joint data center of Max Planck Society for the Advancement of Science (MPG) and University of G\"{o}ttingen.

\appendix
\section{Thermal average cross section}
The thermal average cross-section for the process $W_{D}W_{D} \rightarrow h_{2}h_{2}$ can be expressed as,
\begin{align}
    \langle\sigma v \rangle_{W_{D} W_{D} \rightarrow h_{2} h_{2} } = a_{W_D} + \frac{3b_{W_D}}{x}
\label{sigvWDWD}
\end{align}
where $a_{W_D}$ and $b_{W_D}$ take the following form \cite{Khan:2024biq},
\begin{align}
    a_{W_D} &= \frac{11 g_D^4 \sqrt{1 - r_1}}{288 \pi m_{W_D}^2}
    \left[
    \frac{1-\frac{20}{11}r_1+\frac{15}{11}r_1^2-\frac{5}{11}r_1^3+\frac{1}{16}r_1^4}{(1-\frac{3}{4}r_1+\frac{1}{8}r_1^2)^2}
    \right]
    \,,\nonumber\\
    b_{W_D} &= \frac{13 g_D^4 r_1 \sqrt{1 - r_1}}{1728 \pi m_{W_D}^2}\left[
    \frac{1 - \frac{19}{13}r_1 + \frac{22}{13}r_1^2 - \frac{17}{13}r_1^3 + \frac{127}{208}r_1^4 - \frac{17}{104}r_1^5 + \frac{1}{52}r_1^6}{(1-\frac{1}{2}r_1)^4(1-\frac{1}{4}r_1)^3}
    \right],
    \label{ab-psi-wd}
\end{align}

where $r_1 \equiv (m_{h_2}/m_{W_D})^2$

\section{Decay Width of $W_D$ to different channels}
\label{WD-dec-expression}
The DM matter $W_D$ can have the following decay modes when it has kinetic mixing with the hypercharge gauge boson in the following way,
\begin{itemize}
\item {\bf $W_{D} \rightarrow WW$:} The decay width takes the following form,
\begin{align}
\Gamma_{W_{D} \rightarrow WW} = \frac{e^{2} c^4_{w} M^5_{W_D}}{192 \pi M^4_{w}}
\left( 1 - \frac{4 M^2_{W}}{M^2_{W_D}} \right)^{3/2} \left( 1 + \frac{20 M^2_{W}}{M^2_{W_D}} + \frac{12 M^4_{W}}{M^4_{W_D}} \right) \left( \alpha - \frac{\alpha \hat{M}^2_{W_D} }{\hat{M}^2_{W_D} - \hat{M}^2_{Z} } \right)^2  
\end{align}
In our study, we have considered $\alpha$ to be a very small number; therefore, for our computations, we use $\hat{M} \simeq M$ as shown in Eq.~\ref{mass-releation}.

\item {\bf $W_{D} \rightarrow e^{+}e^{-}$:} The decay width of $W_D$ to the leptons can be expressed as,
\begin{align}
\Gamma_{W_{D} \rightarrow e^{+}e^{-}} &= \frac{e^{2} c^2_{w}}{96 \pi M_{W_D}}
\biggl[ M^2_{e} \left( 16 \alpha^2 - \frac{ 8 \alpha^2 \hat{M}^2_{W_D}}{\hat{M}^2_{W_D} - \hat{M}^2_{Z} } -  \left( \frac{ \alpha \hat{M}^2_{W_D}}{\hat{M}^2_{W_D} - \hat{M}^2_{Z} } \right)^{2} \right) \nonumber \\
& + M^2_{W_D} \left( 8 \alpha^2 - \frac{ 4 \alpha^2 \hat{M}^2_{W_D}}{\hat{M}^2_{W_D} - \hat{M}^2_{Z} } +  \left( \frac{ \alpha \hat{M}^2_{W_D}}{\hat{M}^2_{W_D} - \hat{M}^2_{Z} } \right)^{2} \right)
 \biggr]
 \end{align}

\item {\bf $W_{D} \rightarrow \nu_e \nu_e:$} The decay width of $W_D$ to the neutrinos can be expressed as,
\begin{eqnarray}
\Gamma_{W_{D} \rightarrow \nu_e \nu_e} = \frac{e^{2} c^2_{w} M_{W_D}}{96 \pi} \left(  
 \frac{\alpha \hat{M}^2_{W_D} }{\hat{M}^2_{W_D} - \hat{M}^2_{Z}} \right)^{2}
\end{eqnarray}

\item {\bf $W_{D} \rightarrow \bar u u:$} The decay width of $W_D$ to up-quark can be expressed as, 
\begin{eqnarray}
&& \Gamma_{W_{D} \rightarrow \bar u u} =  \frac{e^{2} M^2_{W_D}}{288 \pi c^2_{w} M^3_{W_D} }
\left( 1 - \frac{4 M^2_{u}}{M^2_{W_D}} \right)^{1/2}
\biggl[ c^4_{w} \biggl( M^2_{u} \left( 64 \alpha^2 - 48 \kappa \alpha^2 - 9 \kappa^2 \alpha^2 \right) \nonumber \\
&&+ M^2_{W_D} \left(32 \alpha^2 - 24 \kappa \alpha^2 + 9 \kappa^2 \alpha^2 \right)
+ \kappa^2 \alpha^2 s^4_{w} \left( 7 M^2_{u} + 17 M^2_{W_D} \right)
 \biggr)
\nonumber \\
&& + 2 c^2_w s^2_w \kappa \alpha \biggl( M^2_{u} \left( 40 \alpha - 33 \kappa\alpha \right)
+ M^2_{W_D} \left( 20 \alpha - 3 \kappa \alpha \right) \biggr) 
  \biggr]
\end{eqnarray}
where $\kappa = \frac{\hat{M}^2_{W_D}}{ \hat{M}^2_{W_D} - \hat{M}^2_{Z}}$.

\item {\bf $W_{D} \rightarrow h_{1} Z$:} Decay width takes the following form,
\begin{eqnarray}
&&\Gamma_{W_{D} \rightarrow h_{1} Z} = \frac{e^{4} M_{W_D} \kappa^{2} \alpha^2 v^{2} \cos^{2} \theta }{768 \pi M^2_{Z} c^4_{w} s^2_{w} } \sqrt{\biggl( 1- \left(\frac{M_{h_1} -M_{Z}}{M_{W_D}} \right)^2 \biggr) \biggl( 1- \left(\frac{M_{h_1} +M_{Z}}{M_{W_D}} \right)^2 \biggr)} \nonumber \\
&&\times \biggl[ 1 + \left( \frac{M_{h_1}}{M_{W_D}} \right)^{4} + 10 \left( \frac{M_{Z}}{M_{W_D}} \right)^{2} + \left( \frac{M_{Z}}{M_{W_D}} \right)^{4} 
- 2 \left( \frac{M_{h_1}}{M_{W_D}} \right)^{2} \biggl(1 + \left( \frac{M_{Z}}{M_{W_D}} \right)^{2} \biggr)
\biggr]
\end{eqnarray}
where $\kappa$ is defined in the previous expression.
\end{itemize}

\end{document}